\documentclass[twocolumn,superscriptaddress,prb,aps,preprintnumbers,longbibliography]{revtex4-2}
\usepackage{amsmath,amssymb,bm,graphicx,color,gensymb,bbold,appendix,xspace}
\usepackage[colorlinks=true, citecolor={blue!80!black}, urlcolor={blue!50!black}, linkcolor = {blue!80!black}]{hyperref}
\usepackage[utf8x]{inputenc}
\usepackage{ulem}
\usepackage{dcolumn}
\usepackage{epsfig}
\usepackage{physics}
\usepackage{xcolor}
\usepackage{comment}

\newcommand{\black}[1]{{\color{black} #1}}

\newcommand{\KSeCo}{K$_2$Co(SeO$_3$)$_2$\xspace}
\newcommand{\NaBaCoP}{Na$_2$BaCo(PO$_4$)$_2$\xspace}

\newcommand{\be}{\begin{equation} }
	\newcommand{\ee}{\end{equation} }
\newcommand{\bea}{\begin{eqnarray} }
	\newcommand{\eea}{\end{eqnarray} }

\newcommand{\maxflip}{
\includegraphics[width=0.4 cm]{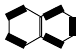}}
\newcommand{\maxfliptwo}{
\includegraphics[width=0.4 cm]{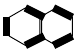}}
\newcommand{\doublehex}{\includegraphics[width=0.4 cm]{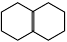}}

\begin{document}

\title{ Wannier states and spin supersolid physics in the triangular antiferromagnet \KSeCo}

\author{M.~Zhu}
\thanks{These two authors contributed equally}
\address{Laboratory for Solid State Physics, ETH Z\"{u}rich, 8093 Z\"{u}rich, Switzerland}

\author{Leandro M. Chinellato}
\thanks{These two authors contributed equally}  \thanks{\\Corresponding author:~\url{lchinell@vols.utk.edu}}
\address{Department of Physics, The University of Tennessee, Knoxville, Tennessee 37996, USA}

\author{V. Romerio}
\address{Laboratory for Solid State Physics, ETH Z\"{u}rich, 8093 Z\"{u}rich, Switzerland}

\author{N. Murai}
\address{J-PARC Center, Japan Atomic Energy Agency, Tokai, Ibaraki 319-1195, Japan}

\author{S. Ohira-Kawamura}	
\address{J-PARC Center, Japan Atomic Energy Agency, Tokai, Ibaraki 319-1195, Japan}

\author{Christian Balz}
\address{ISIS Neutron and Muon Source, STFC Rutherford Appleton Laboratory, Didcot OX11 0QX, United Kingdom}
\address{Neutron Scattering Division, Oak Ridge National Laboratory, Oak Ridge, Tennessee 37831, USA}

\author{Z. Yan}
\address{Laboratory for Solid State Physics, ETH Z\"{u}rich, 8093 Z\"{u}rich, Switzerland}
\author{S. Gvasaliya}
\address{Laboratory for Solid State Physics, ETH Z\"{u}rich, 8093 Z\"{u}rich, Switzerland}

\author{Yasuyuki Kato}
\address{Department of Applied Physics, University of Fukui, Fukui 910-8507, Japan}

\author{C. D.~Batista}
\address{Department of Physics, The University of Tennessee, Knoxville, Tennessee 37996, USA}
\address{Quantum Condensed Matter Division and Shull-Wollan Center, Oak Ridge National Laboratory, Oak Ridge, Tennessee 37831, USA}

\author{A.~Zheludev}
\address{Laboratory for Solid State Physics, ETH Z\"{u}rich, 8093 Z\"{u}rich, Switzerland}

\begin{abstract}
    We combine ultra-high-resolution inelastic neutron scattering and quantum Monte Carlo simulations to study thermodynamics and spin excitations in the spin-supersolid phase of the triangular lattice XXZ antiferromagnet \KSeCo under zero and non-zero magnetic field. BKT transitions signaling the onset of Ising and supersolid order are clearly identified, and the Wannier entropy is experimentally recovered just above the supersolid phase. At low temperatures,  with an experimental resolution of about 23 $\mu$eV, no discrete coherent magnon modes are resolved within a broad scattering continuum. Alongside gapless excitations, a pseudo-Goldstone mode with a 0.06 meV gap is observed. A second, higher-energy continuum replaces single-spin-flip excitations of the Ising model. Under applied fields, the continuum evolves into coherent spin waves, with Goldstone and pseudo-Goldstone sectors responding differently. The experiments and simulations show excellent quantitative agreement.
\end{abstract}

\date{\today}
\maketitle
\section{Introduction}

Identifying and understanding novel quantum phases of matter has been a central topic in modern condensed matter physics. 
The search for a supersolid, an exotic quantum state which is simultaneously a solid and a superfluid \cite{Boninsegni2012_RMP}, is one of the most celebrated examples among such efforts. Originally proposed theoretically in solid helium~\cite{Leggett1970,Andreev1971}, the supersolid phase has since been studied extensively across various systems,  including cold atoms \cite{Li2017,Tanzi2019,Guo2019}, hard-core bosons \cite{Heidarian2005,Wessel2005, Melko2005}, and quantum spin systems \cite{Tsutomu2000, Sengupta2007,Heydarinasab2018,Gao2022, Wang2023}. Despite significant theoretical progress, experimental identification of a supersolid phase remains a challenge.

 The quantum $S$ = 1/2 antiferromagnetic XXZ spin Hamiltonian on a triangular lattice \cite{Miyashita1986, Yamamoto2014, SellmannZhangEggert_PRB_2015_AnotherXXZtriangularPhD,Bordelon2019,Wu2022a,Ma2016} presents a promising platform for realizing a supersolid state~\cite{Sengupta2007,Schmidt08}. The ``parent'' Ising model has been famously solved by Wannier \cite{wannier1950}. The ground state is macroscopically degenerate with no magnetic order and a residual entropy of 0.323$R$. Members of this manifold are states where each triangle of the lattice carries two spins opposed to the third one. Since the model is classical, excitations are flat bands corresponding to flips of single spins. 
An ordered state is formed when this degeneracy is lifted  by off-axial interactions in the XXZ model. The supersolid interpretation is based a one-to-one mapping between a spin-$1/2$ magnet and a gas of hard-core bosons on the lattice~ \cite{Matsubara56}. In this mapping, the local magnetization along the $z$-axis, $m^z_j$, corresponds to the local boson density, while the magnitude and direction of the $xy$ component of the local magnetization, $m^{xy}_j$, determine the amplitude and phase of the local Bose-Einstein condensate (BEC) order parameter. The ``solid'' spin ordering is associated with a collinear spin density wave along the $z$-axis, while the ``BEC'' translates into some form of $XY$-ordering.

The superfluid current of a bosonic gas in the BEC state corresponds to a superfluid spin current in the pure XXZ model. However, it is important to note that the spin Hamiltonian of a real spin system always includes terms that break the continuous symmetry associated with the conservation of any component of the net magnetization. Consequently, real spin systems lack a continuity equation, resulting in short-lived spin currents and making the concept of a spin superfluid poorly defined. Nevertheless, we will continue to use the term ``spin supersolid'' to describe spin systems that are well approximated by U(1)-invariant models, which undergo a transition into a ``solid'' phase at a critical temperature $T_{c1}$, followed by a second transition into a ``superfluid'' phase that coexists with the solid at a lower temperature $T_{c2}$. As we will see here, a key aspect that remains robust in the spin incarnation of this phenomenon is the strong quantum character of the low-energy excitations.

\begin{figure}[t!]
    \centering
    \includegraphics[width=0.99\columnwidth]{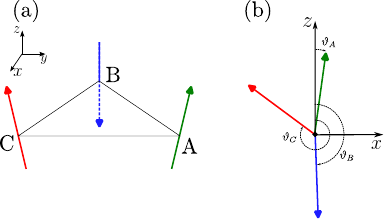}
    \caption{\textbf{Classical ground state magnetic structures.} (a) Classical Y-phase. The order breaks the U(1) symmetry of the Hamiltonian selecting one orientation on the $xy$-plane and the ${\text C}_3$ lattice geometry selecting one sublattice for the spin down. (b) The classical ground state exhibits accidental continuous degeneracy due to the freedom to select one of the three spin angles, with the other two constrained by the derived relation Eq. \ref{Eq:ang_rel}.}
    \label{Fig:Y}
\end{figure}

Various theoretical approaches have predicted a spin supersolid ground state for easy-axis exchange anisotropy ($0 < \alpha = J_{xy}/J_{zz} < 1$) \cite{Wang09, Heidarian2010, Gao2022, Yamamoto2014, gallegos2024}. The magnetic ordering exhibits a three-sublattice planar ``Y''-spin structure, consisting of a commensurate spin density wave in the out-of-plane $z$-component with zero net magnetization, alongside an in-plane transverse $xy$ moment. Notably, this ``Y''-spin structure differs from the classical ordering for $\alpha \ll 1$, where the classical ``Y'' [see Figure~\ref{Fig:Y}(a)] structure has a net magnetization close to one-third of the saturation value. In contrast, quantum fluctuations significantly reduce the length of the two quasi-parallel spins to nearly half the length of the spin that is nearly anti-parallel to the other two. This strong quantum effect, driven by the macroscopic ground state degeneracy of the pure Ising model, suggests the possibility of highly non-classical spin dynamics. 
The system undergoes a quantum phase transition to a collinear up-up-down (uud) phase in a magnetic field ($m/m_{\rm sat} = 1/3$), whose spin dynamics is expected to be well described by a semi-classical theory~\cite{Alicea09,Kamiya18}.

While the spin supersolid ground state is well understood theoretically~\cite{gallegos2024,Yamamoto2014,Wang2009,Melko2005,Jiang2009,Xu2025,Heidarian2005, Murthy1997}, the spectrum of magnetic excitations remains an open question. Recent inelastic neutron scattering (INS) study on the triangular lattice easy-axis XXZ antiferromagnet \KSeCo \cite{Zhu2024} sheds new light on this issue. The ground state of \KSeCo has been proposed as a spin supersolid phase with a ``Y'' structure \cite{Zhu2024, Chen2024, Zhong2020}. Interestingly, the spin dynamics is highly unusual, with the low-energy magnetic spectrum dominated by a broad continuum of excitations, aside from the gapless Goldstone modes, and a roton-like excitation dip observed at the $\textrm{M}$ points of the Brillouin zone.

In stark contrast to the zero-field spectrum, the magnetic excitations in the field-induced uud plateau phase become sharp, resolution-limited magnons, well described by semiclassical spin wave theory \cite{Zhu2024}. Since excitation continua in frustrated magnets are often seen as indicators of proximity to a quantum spin liquid state \cite{Anderson1973, Balents2010, Savary2016, Broholm2020, Ghioldi2022}, understanding the origin of this continuum in \KSeCo is crucial for exploring quantum dynamics scenarios without a semiclassical counterpart.

In this article, we address this question by combining new ultra-high-resolution inelastic neutron scattering and additional thermodynamic measurements with numerical quantum Monte Carlo (QMC) simulations of a sign-problem-free low-energy model.
In zero field, both in thermodynamics and in spectroscopy,   we identify the energy scales that separate the ``Wannier states'' responsible for the formation of the supersolid state from high-energy excitations consisting of plaquettes with three parallel spins.
We study the low-energy spectrum in the Wannier subspace $\mathcal{W}$ to within an energy resolution of approximately 23~$\mu$eV, putting experimental bounds on intensities and energies of any coherent excitations relative to the continuum.  At the $\textrm{K}$-point, in addition to gapless excitations, we find an additional pseudo-Goldstone mode with a tiny gap of 60~$\mu$eV. At higher energies of 3~meV, where flat spin flip modes 
are to be found in the Ising model, we observe a second structured and dispersive continuum. 
In applied fields, we follow the evolution of the low-energy excitations through the supersolid state and all the way to the uud phase boundary. The continuum coalesces into what eventually becomes sharp spin waves. The pseudo-Goldstone gap at the $\textrm{K}$ point increases with magnetic field, even as the roton-like dip at the $\textrm{M}$ point stays prominent. In contrast, the gapless excitations remain gapless until the quantum critical point, but their roton minimum quickly vanishes.

\begin{figure}[!t]
    \centering
    \includegraphics[width=\columnwidth]{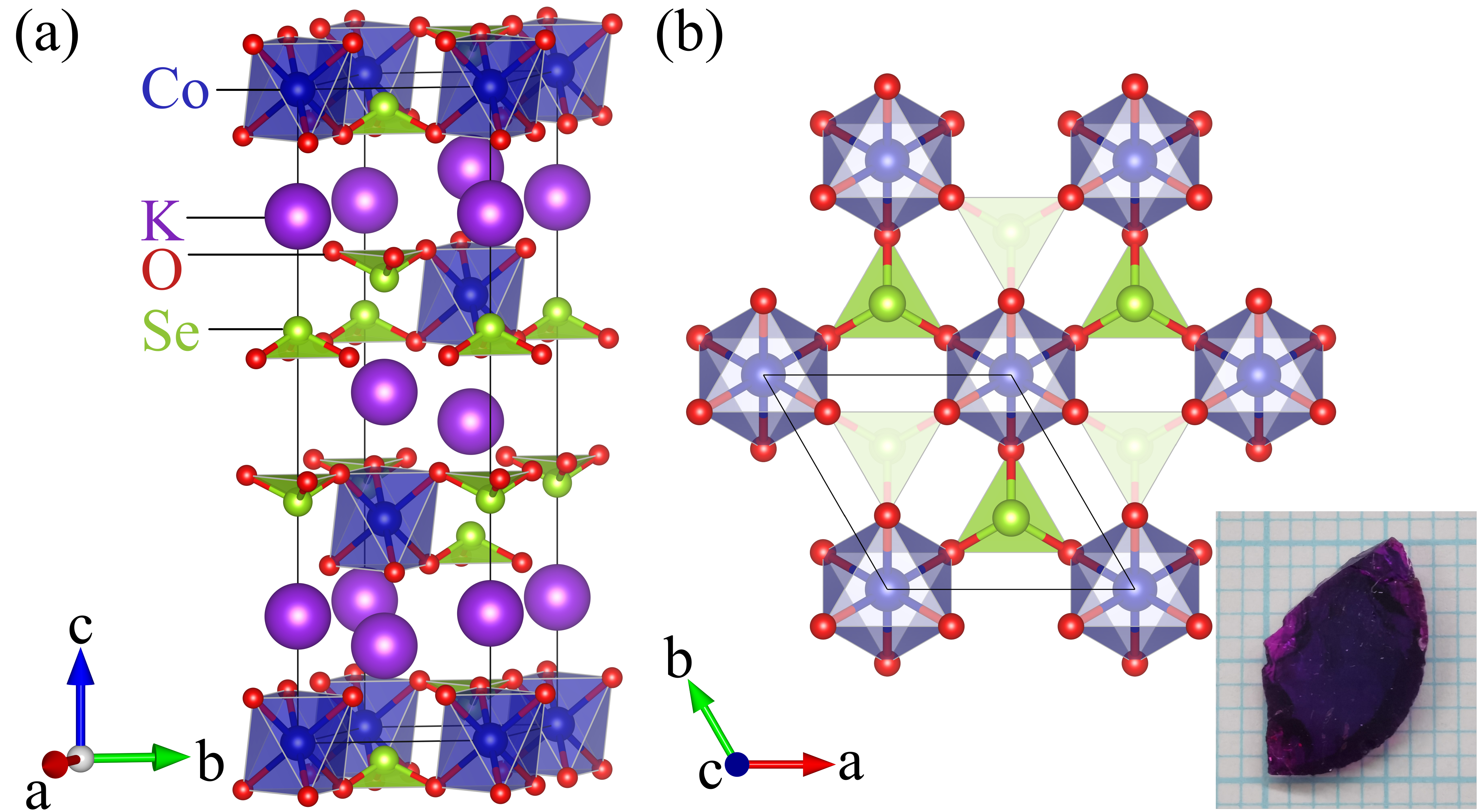}

    \caption{\textbf{Crystal structure of \KSeCo.} A schematic view of the crystal structure of $\text{K}_2\text{Co}(\text{Se}\text{O}_3)_2$. (a) Unit cell with ABC-stacked triangular planes of Co$^{2+}$ ions. 
    (b) Top view of a single triangular plane. Inset: photo of a $\text{K}_2\text{Co}(\text{Se}\text{O}_3)_2$ single crystal on a millimeter grid paper.}
    
    \label{fig:crys_str}
\end{figure}

\begin{figure}[!t]
    \centering   
    \includegraphics[width=1.0\columnwidth]{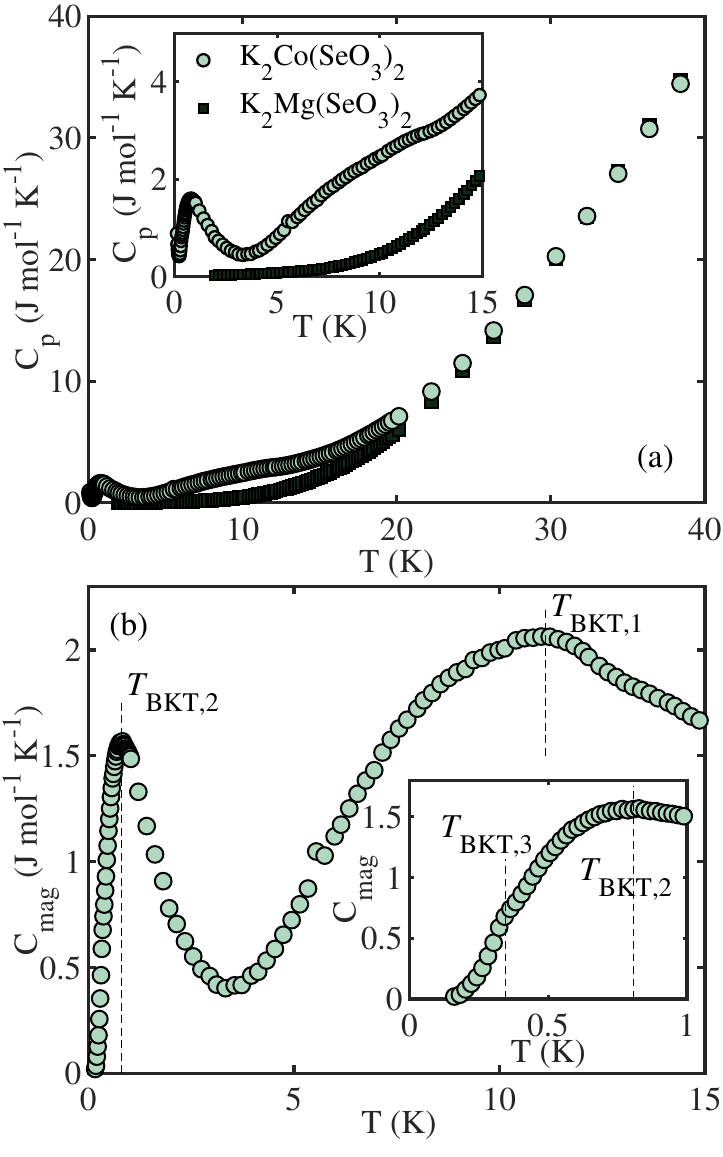}
    \caption{\textbf{Temperature dependence of specific heat in \KSeCo.} (a)  Heat capacity measured in~\KSeCo (shaded circles) and in the non-magnetic counterpart K$_2$Mg(SeO$_3$)$_2$ (solid squares) as a function of temperature. A 1.08 scale factor was applied to the latter, to match the heat capacities of the two materials in the range 50 to 100 K. (b) Temperature dependence of the magnetic contribution to specific heat in \KSeCo. The phonon contribution was estimated based on the scaled K$_2$Mg(SeO$_3$)$_2$ data. The two prominent peaks correspond to the two higher-temperature BKT transitions. Inset: expanded view of the low-temperature magnetic heat capacity. A weaker feature at lower temperature marks the third transition. Actual BKT transition temperatures are typically lower than the corresponding specific heat maximum \cite{Sengupta03}.}
    \label{Fig:high_T_cv}
\end{figure}

\begin{figure}[t!]
     \centering
     \includegraphics[width=\columnwidth]{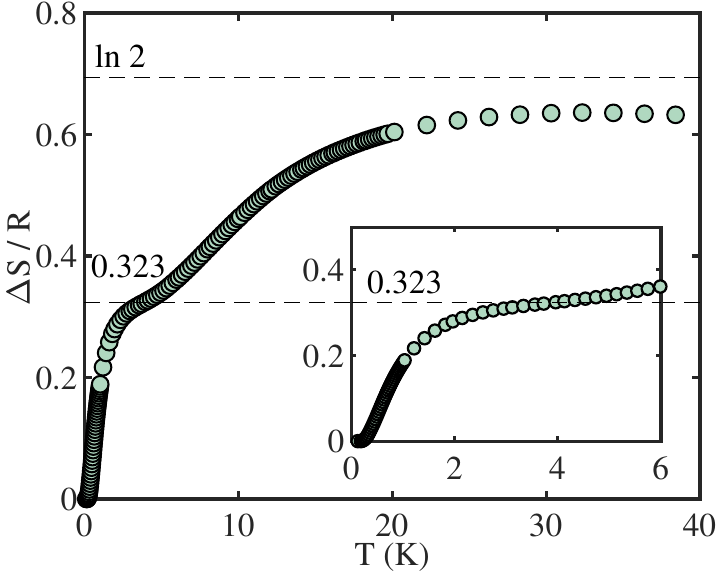}
     \caption{\textbf{Magnetic entropy as a function of temperature in \KSeCo.} Temperature dependence of magnetic entropy in  $\text{K}_2\text{Co}(\text{Se}\text{O}_3)_2$ in zero applied field, obtained by integrating the measured magnetic specific heat $C_{\text{mag}}/T$. The dashed lines indicate the Wannier entropy and the expected high-temperature value $\ln{2}$. The inset is a zoomed-in view to highlight the plateau at the Wannier entropy. The 10\% discrepancy between the high-temperature data and the expected saturation value is attributed to imperfect phonon subtraction. This effect is expected to be much smaller at temperatures below 10~K.}
     \label{Fig:Ent_exp}
 \end{figure}

\section{Results}

\subsection{Material}
\label{Sec:Mat}

\KSeCo is an easy-axis $S=1/2$ triangular lattice antiferromagnet [see Figure~\ref{fig:crys_str}] very close to the Ising limit with $J_{zz}$ = 3.1 meV, $\alpha = 0.08$~\cite{Zhu2024} deduced from high-field magnetometry measurements. At zero magnetic field, it exhibits no long-range magnetic order down to $T = $ 0.35 K \cite{Zhong2020}. Below that our calorimetric study revealed an additional low-temperature phase \cite{Zhu2024}, which has been proposed to be the theoretically predicted spin-supersolid \cite{Heidarian2010,Yamamoto2014}. When a magnetic field is applied along the easy axis, a quantum critical point is found at $\mu_0 H_c = 0.8 (1)$ T, beyond which the system transitions into the uud plateau phase. Neutron diffraction studies have revealed quasi-two-dimensional magnetic reflections with propagation vectors (1/3, 1/3) in both the Y- and the uud phases, suggesting magnetic structures with a $\sqrt{3} \times \sqrt{3}$ supercell qualitatively in agreement with the expected three-sublattice order in the spin-supersolid and spin-solid phase  \cite{Zhu2024}.

\subsection{Specific heat and entropy}
\label{Sec:specficheat}

Figure~\ref{Fig:high_T_cv}(a) shows the zero-field specific heat of \KSeCo measured up to high temperatures (open circles). The phonon contribution is approximated by the measured specific heat of the isostructural nonmagnetic counterpart K$_2$Mg(SeO$_3$)$_2$ (solid squares). An empirical scale factor of 1.08 was applied in order to match the heat capacities of the two systems in the temperature range 50$-$100 K. The upturn below 0.2 K \cite{YingFu2025} is ascribed to a nuclear Schottky-like feature (see Supplementary Note 1). The magnetic heat capacity $C_{\text{mag}}$ of \KSeCo is plotted as a function of temperature 
in Figure~\ref{Fig:high_T_cv}(b). Here the phonon and the nuclear Schottky contributions at the lowest temperatures were subtracted from the total heat capacity (see Supplementary Figure 1). Signatures of three distinct phase transitions are observed at $T_{\text{BKT,1}} \approx$ 11 K, $T_{\text{BKT,2}} \approx$ 0.8 K, and $T_{\text{BKT,3}} \approx$ 0.35 K (dashed lines). While the third transition is somewhat weakly defined, its position exactly corresponds to the zero-field asymptotic of a much sharper lambda-anomaly in applied field \cite{Zhu2024}. As to be discussed in more detail in Sec.~\ref{sec:theory}, these features are BKT transitions corresponding to the Ising and XY orderings of the supersolid phase, respectively.

In Figure~\ref{Fig:Ent_exp} we show the magnetic entropy $\Delta S(T)$ of \KSeCo\ as a function of temperature, obtained by integrating the magnetic heat capacity $C_{\text{mag}}/T$ from the lowest temperature measured. A plateau is observed at $ T_{\rm cr} \simeq 2.5~\text{K}$. This crossover temperature is of similar order of magnitude as the exchange constant for the in-plane spin component $J_{xy}/k_B$, which suggests a separation of energy scales in the magnetic excitation spectrum. The plateau value, $\Delta S/R \approx$ 0.323, aligns well with the zero-point entropy of the pure Ising model calculated by Wannier. This implies that the low-energy excitations arise from the ``Wannier states".

\begin{figure}[t!]
	    \includegraphics[width=1.0\columnwidth]{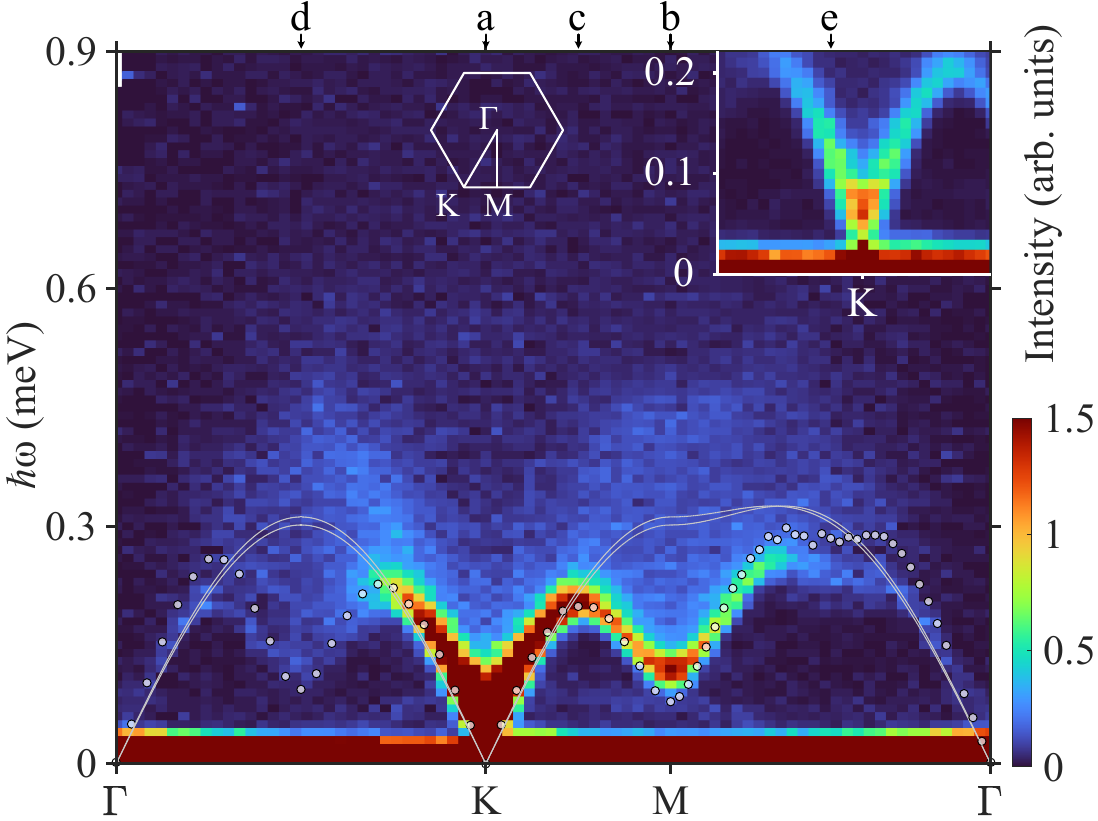}

	\caption{\textbf{High-resolution magnetic excitation spectrum of \KSeCo in the supersolid phase.} False-color plot of inelastic neutron scattering intensity measured in \KSeCo in zero applied field and $T=0.07$~K as a function of energy and wave vector transfers along high-symmetry directions on the AMATERAS spectrometer. The incident neutron energy is $E_i$ = 1.69~meV. The energy resolution is 23 $\mu$eV at the elastic line. This is to be compared to the 36~$\mu$eV resolution of the experiments described in Ref.~\cite{Zhu2024}. The data are integrated over $\pm 0.5$ r.l.u. in $l$ and $\pm 0.05$ r.l.u. in the $(h,k,0)$ plane perpendicular to the cut, and averaged over equivalent paths shown in Supplementary Figure 3.
    ~No background has been subtracted. Both longitudinal and transverse spin fluctuations contribute to scattering, the intensity being proportional to $S^{zz}(\mathbf{q},\omega)+S^{\bot}(\mathbf{q},\omega)$. White solid lines are the dispersion curves calculated by linear spin wave theory. Circles are first poles of $S^{zz}(\mathbf{q},\omega)$ calculated by QMC. The white hexagon represents the Brillouin zone boundary. $\Gamma$, $\textrm{K}$ and $\textrm{M}$ label high symmetry points in the reciprocal space. Inset: a zoom-in view of the low-energy spectrum near the $\textrm{K}$ point with a different color scale to highlight the gapped pseudo-Goldstone excitation. Labels on the top axis indicate the position of energy cuts shown in Figure~\ref{AMA_E_cuts}.  }
 \label{AMA_high_resolution}
\end{figure}

\subsection{Inelastic neutron scattering}
\label{Sec:ins}

The separation of energy scales in spin dynamics is clearly evident in our inelastic neutron scattering measurements.  
In Figure~\ref{AMA_high_resolution} we show the low-energy magnetic excitation spectrum of \KSeCo up to 0.9 meV measured by high-resolution inelastic neutron scattering at AMATERAS as a function of energy and high-symmetry directions in the reciprocal space at $T =  70 ~\textrm{mK}$ and zero magnetic field. The incident neutron energy is $E_i = 1.69$ meV, giving a resolution of 23 $\mu$eV at elastic scattering (to be compared to the resolution $\Delta E \approx$ 36 $\mu$eV for the data reported in \cite{Zhu2024}). The spectrum is dominated by the broad, gapless continuum of excitations, deviating drastically from the predictions of the linear spin wave theory (magnon dispersion in solid lines), and no sharp, coherent magnon modes can be resolved even with the improved energy resolution. Note that in principle,  both longitudinal and transverse fluctuations, $S^{zz}$ and $S^{\perp}$, contribute to the scattered intensities, as the measurements were performed with $l \approx 0$. However, due to the highly anisotropic g-tensor (a crude estimate based on the Curie Weiss fit of the low-temperature magnetic susceptibility suggests g$_{zz}$/g$_{\perp} \approx 4$), most of the scattering intensity likely stems from the longitudinal channel $S^{zz}(\mathbf{q},\omega)$.

\begin{figure}[!t]
    \centering
    \includegraphics[width=0.8\columnwidth]{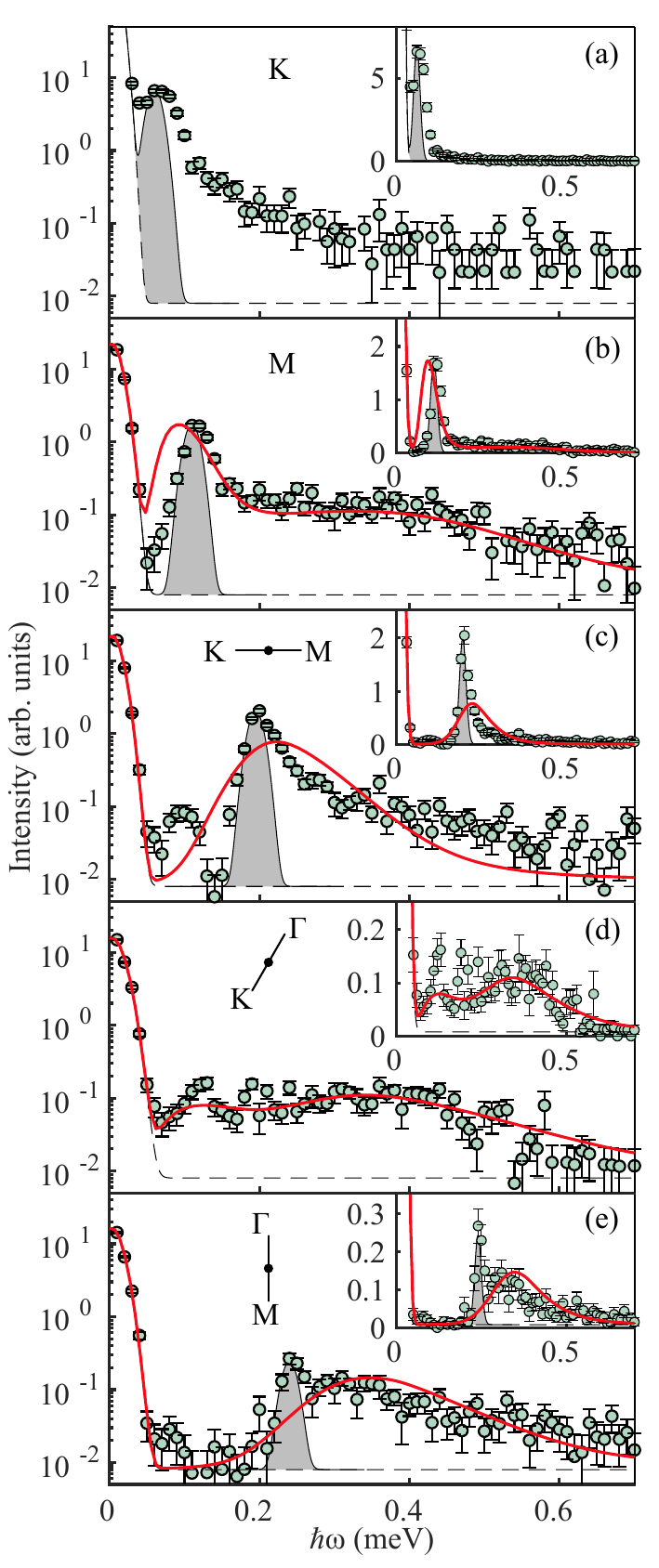}
    \caption{\textbf{Energy cuts of the magnetic excitation spectrum in \KSeCo in the supersolid phase.} Energy cuts through the INS data shown in Figure~\ref{AMA_high_resolution} at representative wave vectors as indicated.  Both longitudinal and transverse spin fluctuations contribute to scattering, the intensity being proportional to $S^{zz}(\mathbf{q},\omega)+S^{\bot}(\mathbf{q},\omega)$. (a)  $\textrm{K}$ point, (b) $\textrm{M}$ point, (c) half way between $\textrm{K}$ and $\textrm{M}$, (d) half way between $\Gamma$ and $\textrm{K}$, and (e) half way between $\textrm{M}$ and $\Gamma$. The solid lines represent QMC calculations for  $S^{zz}(\mathbf{q},\omega)$ only. An arbitrary global scale factor is applied to match the intensity of the $\textrm{M}$ point. The measured spectra have been averaged over equivalent wave vectors accessible in the experiment. Note the y-axis is in logarithmic scale. Insets show the same plot with y-axis in linear scale. Shaded Gaussians represent the energy resolution. The dashed lines are fits to a constant background and the elastic scattering peak. }
    \label{AMA_E_cuts}
\end{figure}

The continuum of excitations is also obvious in the energy-dependent cuts plotted in Figure~\ref{AMA_E_cuts} at several representative wave vectors: $\textrm{M}$ point [Figure~\ref{AMA_E_cuts}(b)], half way between the $\textrm{K}$ and $\textrm{M}$ points [Figure~\ref{AMA_E_cuts}(c)], and half way between the $\textrm{M}$ and $\Gamma$ points [Figure~\ref{AMA_E_cuts}(e)]. A sharp excitation feature may be present at the lower boundary of the continuum for these wave vectors, whereas it can not be resolved at the wave vector halfway between the $\Gamma$ and $\textrm{K}$ points [Figure~\ref{AMA_E_cuts}(d)]. Note that the intensity of the scattered neutrons is plotted in the logarithmic scale, and the corresponding plots with linear scale are displayed in the insets. The observation of a sharp feature at the lower boundary of a continuum of excitations in low-dimensional quantum magnets is often interpreted as a bound state of fractionalized spin excitations \cite{Stone2003}. In the present study, we cannot resolve any gap between the sharp excitation and the rest of the continuum within our experimental resolution, which seems to support such a scenario. Nevertheless, one can never exclude the possibility that there exist closely spaced coherent magnon modes where the difference in energy is much smaller than our experimental resolution. Even then, our high-resolution INS data imposed strong constraints on the upper limit of the energy spacing between potential discrete magnon modes. Further, our data allow us to make an accurate estimate on the spectral weight contributions of the sharp magnon modes, if existing, at the lower boundary of the continuum.  At the wave vectors investigated [Figure~\ref{AMA_E_cuts}(b),(c),(e)], the weight of the sharp excitation (represented by a shaded Gaussian in which FWHM corresponds to the energy resolution) is only about 40\%, implying that at least 60\% of the total spectral weight is contributed by the continuum.

 At the $\textrm{K}$ point, in addition to the gapless Goldstone excitations as expected from the spontaneously broken U(1) symmetry in the supersolid phase, a gapped excitation is observed at approximately $E \approx 0.06$ meV [Figure~\ref{AMA_E_cuts}(a)]. This excitation corresponds to a pseudo-Goldstone mode and is prominently visible in the energy-momentum plot of the magnetic excitation spectrum near the $\textrm{K}$ point, as illustrated in the inset of Figure~\ref{AMA_high_resolution}.

\begin{figure}[!h]
	
    \includegraphics[width=\columnwidth]{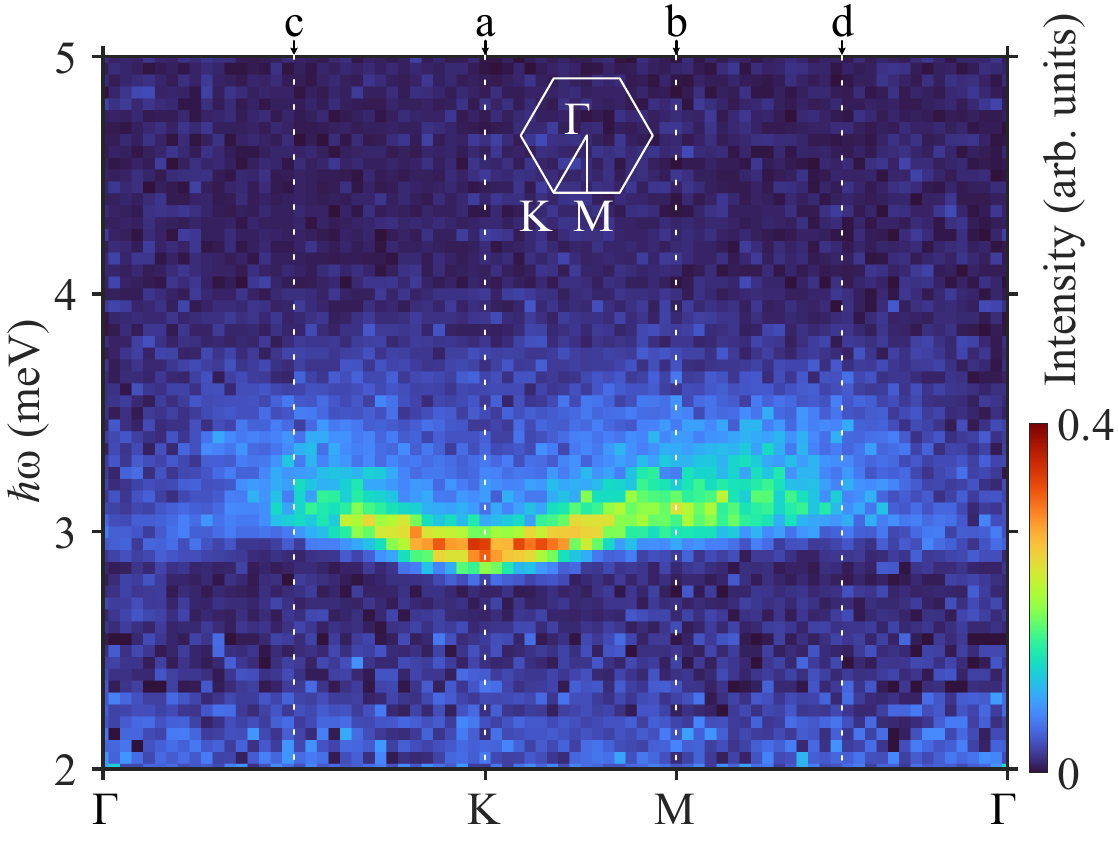}
	\caption{\textbf{High-energy magnetic excitations in \KSeCo in the supersolid phase.} False-color plot of neutron scattering intensity measured on \KSeCo at the AMATERAS spectrometer with $E_i$ = 7.73 meV at $T=70$ mK and $H$ = 0, as a function of energy and wave vector transfers along high-symmetry directions in the Brillouin zone. The energy resolution is 0.24 meV at the elastic position.  The data were integrated over $\pm 0.5$ r.l.u. in $l$ and $\pm 0.05$ r.l.u. in the $(h,k,0)$ plane perpendicular to the cut, and averaged over equivalent paths shown in Supplementary Figure 5.
    ~No background has been subtracted. White hexagon represents the Brillouin zone boundary. $\Gamma$, $\textrm{K}$ and $\textrm{M}$ label high symmetry points in the reciprocal space. Labels above the top axis indicate the positions of energy-cuts shown in Figure~\ref{AMA_3meV_E_cuts}.}
 \label{AMA_3meV}
\end{figure}

\begin{figure}[!t]
    \centering
    \includegraphics[width=1.0\columnwidth]{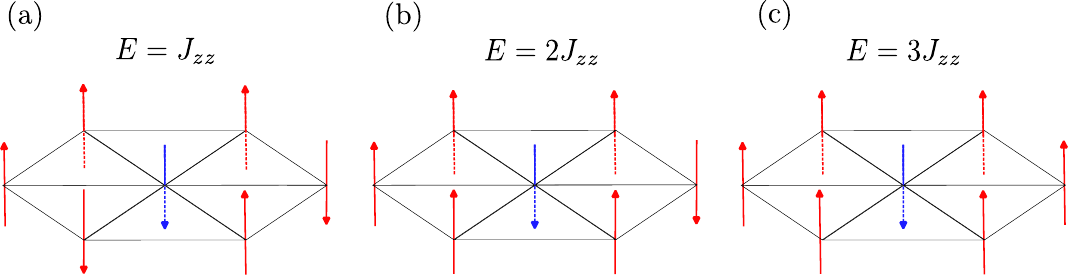}
    \caption{{\bf High energy excitations at zero field.} Sketch of the high-energy excitations in the Ising model at zero field. The excitation corresponds to a flip in the central blue spin changing the net magnetization in $\pm 1$ and the energy in (a) $J_{zz}$; (b) $2J_{zz}$; (c) $3J_{zz}$.}
    \label{Fig:High_ene_exitations}
\end{figure}

\begin{figure}[!h]
	\includegraphics[width=\columnwidth]{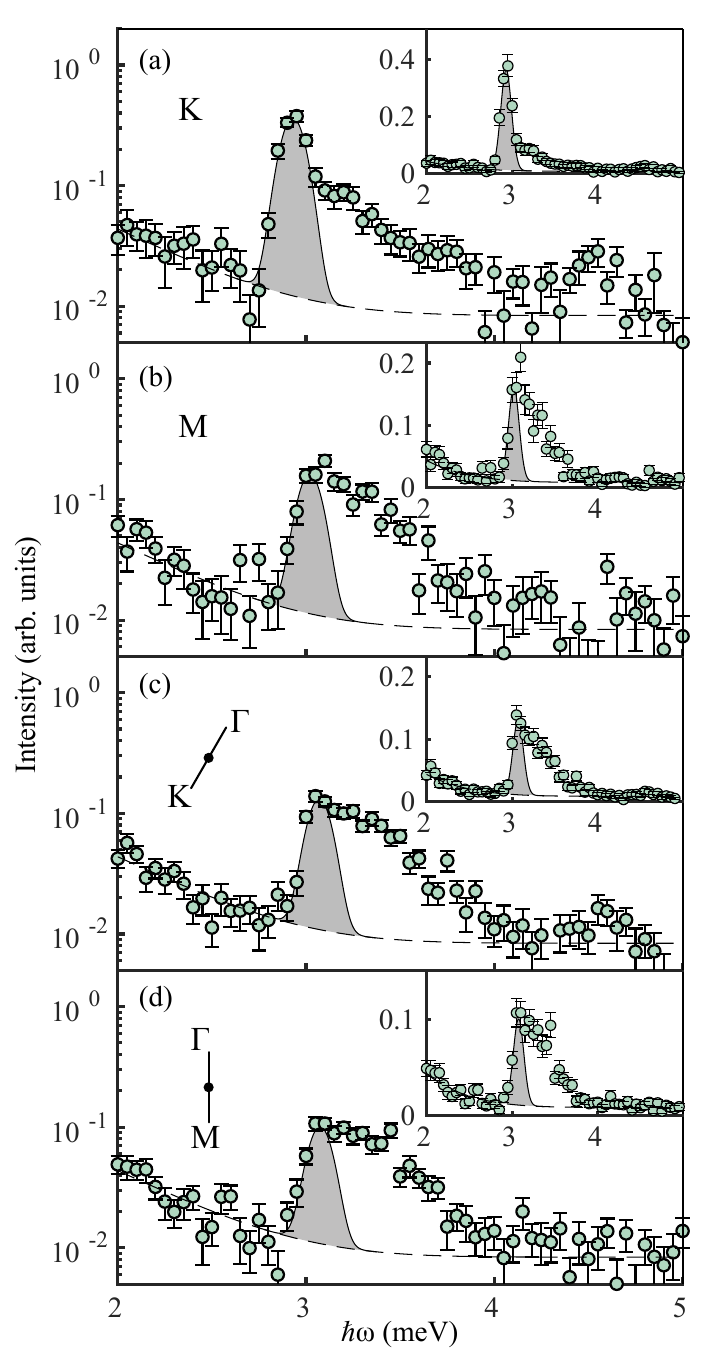}
	\caption{\textbf{Energy cuts of the high-energy magnetic excitations in \KSeCo in the supersolid phase.} Energy-cuts through the INS data shown in Figure~\ref{AMA_3meV} at representative wave vectors: (a)  $\textrm{K}$ point, (b) $\textrm{M}$ point, (c) half way between $\Gamma$ and $\textrm{K}$, and (d) half way between $\textrm{M}$ and $\Gamma$. The wave vectors are indicated in Figure~\ref{AMA_3meV} by arrows. The spectra have been averaged over equivalent wave vectors. Note the y-axis is in logarithmic scale. The inset shows the same plot with $y$-axis on a linear scale. The shaded Gaussian represents the energy resolution. The dashed lines are empirical fits to the background.}
 \label{AMA_3meV_E_cuts}
\end{figure}

Apart from these low-energy excitations, we also observe higher-energy excitations near 3 meV, using incident neutron energy $E_i = 7.73$ meV, as shown in Figure~\ref{AMA_3meV}. In the pure Ising model, dispersionless single-magnon bands are predicted at energies $E = n J_{zz}$ with $n = 1, 2, 3$, corresponding to spin-flip excitations within a locally magnetized environment as illustrated in Figure~\ref{Fig:High_ene_exitations}.
The observed high-energy excitation corresponds to the spin flip excitation with $n = 1$. 
It becomes slightly dispersive due to the nonzero $J_{xy}$ ($\alpha \approx 0.08$ in \KSeCo), with the bottom of the branch located at the $\textrm{K}$ point. The average energy of the excitation is consistent with the exchange interaction $J_{zz} = 3.1$ meV, as deduced from high-field magnetization data \cite{Zhu2024}. Instead of a sharp coherent mode, this excitation manifests as a gapped continuum, with its broadening attributed to strong quantum fluctuations within the Wannier sector. Similar to the lower-energy excitations, a sharp mode might exist at the lower bound of the continuum, as suggested by the shaded Gaussians in the energy-dependent cuts plotted in Figure~\ref{AMA_3meV_E_cuts}.
Note that to enhance the counting statistics of the neutron intensity, the INS data presented in Fig.~\ref{AMA_high_resolution}-\ref{AMA_3meV_E_cuts} have been averaged over equivalent paths in the reciprocal space accessible in the experiment, as shown in Supplementary Figure 3 and Supplementary Figure 5.

\begin{figure*}[ht]
   
     \includegraphics[width=0.9\textwidth]{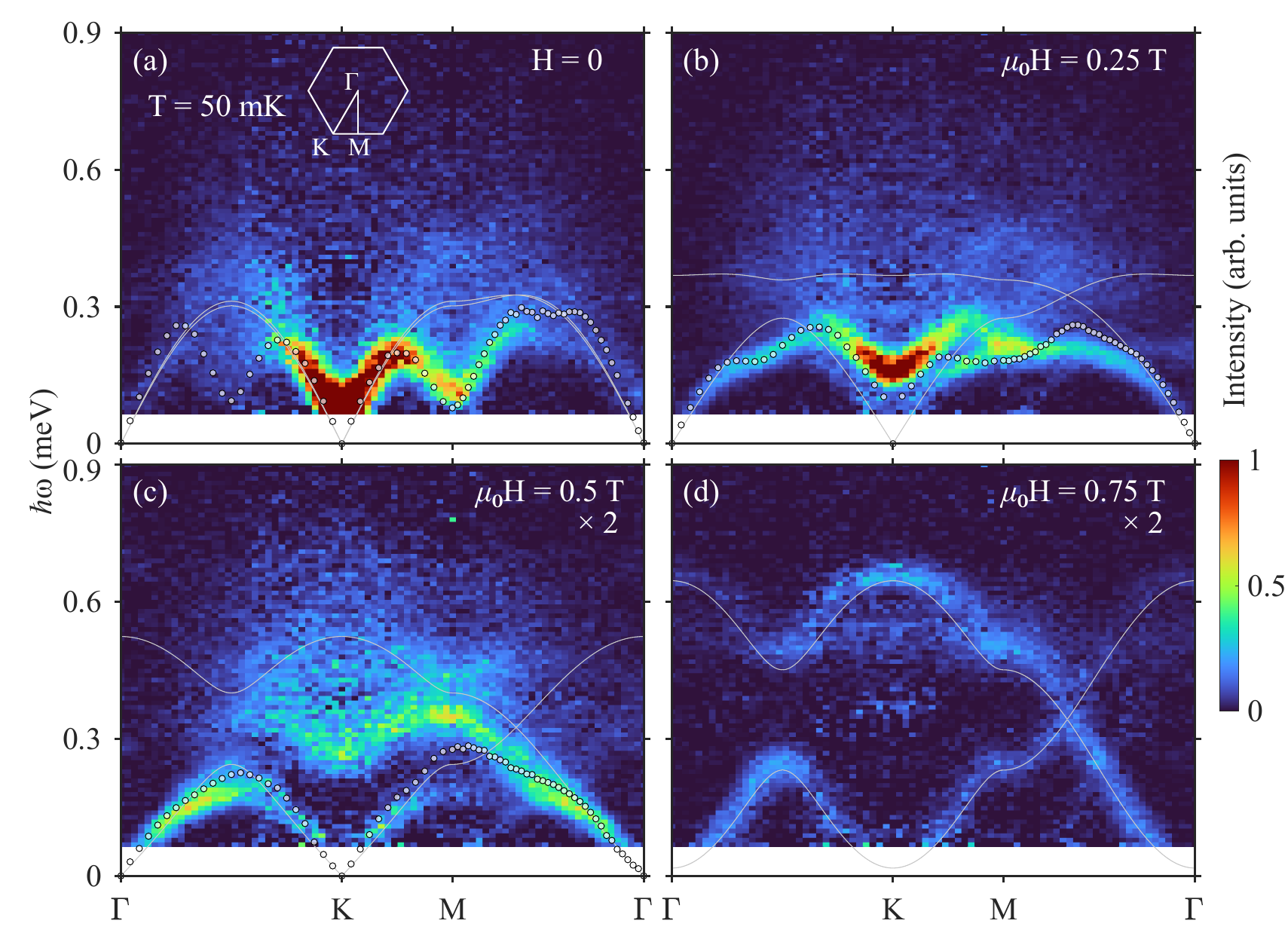}
	\caption{\textbf{Evolution of the magnetic excitation spectrum of \KSeCo in the supersolid phase in a magnetic field. } False-color plots of neutron scattering intensity measured on \KSeCo on the LET spectrometer at $T$ = 50 mK as a function of energy and wave vector along high-symmetry directions of the Brillouin zone in a magnetic field applied along the $c$ axis: (a) $\mu_0H = 0$, (b) 0.25~T, (c) 0.5~T, and (d) 0.75~T. The incident neutron energy is $E_i$ = 2.15~meV, the resolution is 0.04~meV at the elastic line. The data were integrated over $\pm 0.75$ r.l.u. in $l$ and $\pm 0.05$ r.l.u. in the $(h,k,0)$ plane perpendicular to the cut, and averaged over equivalent paths shown in Supplementary Figure 4.
    The spectra measured at 7~T have been subtracted to remove the background from the sample environment. Note that in (d) the residual scattering at the $\textrm{K}$ point near 0.4~meV is due to spurious scattering from the cryomagnet and imperfect background subtraction. Solid lines are the dispersion curves calculated by the linear spin wave theory. Circles are first poles of $S^{zz}(\mathbf{q},\omega)$ calculated by QMC. The white hexagon
represents the Brillouin zone boundary. $\Gamma$, $K$ and $M$ label
high symmetry points in the reciprocal space.}
 \label{spectra_LET}
\end{figure*}

\begin{figure}[ht]
    
      \includegraphics[width=\columnwidth]{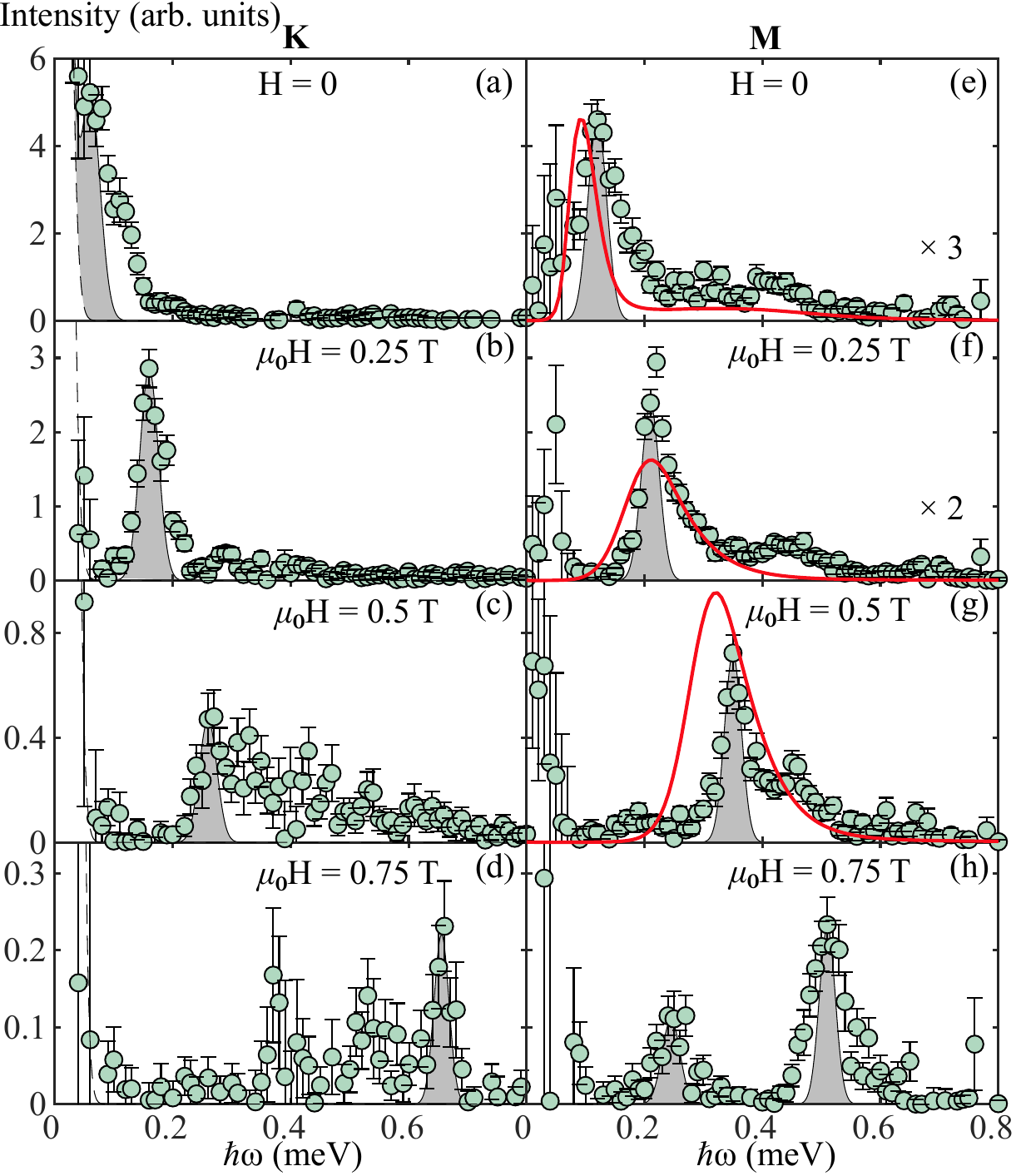}
	\caption{\textbf{Energy cuts of the magnetic excitation spectrum in \KSeCo in a magnetic field.} Magnetic field evolution of the neutron scattering intensity in \KSeCo, as measured at the LET spectrometer with $E_i=2.15$~meV, as a function of energy transfer at the (a-d) K point and (e-h) M point. A magnetic field of $\mu_0H$ = 0, 0.25, 0.5, and 0.75~T, respectively, is applied along the crystallographic $c$ axis. The spectra have been averaged over equivalent wave vectors accessible in the experiment. The spectra measured at 7~T have been subtracted to remove the background from the sample environment. Note in (d) the scattering near 0.4 meV is spurious and due to imperfect backbround subtraction. The scale on the y-axis is linear. The shaded Gaussians represent the calculated energy resolution. The solid lines represent QMC calculations for  $S^{zz}(\mathbf{q},\omega)$ only. An arbitrary global scale factor is applied to match the intensity of the $\textrm{M}$ point at $H$ = 0. The dashed lines are fits to the elastic peak.}
 \label{spectra_LET_E_cut}
\end{figure}

To better understand the nature of the low-energy continuum of excitations in the supersolid phase of \KSeCo, we have investigated their evolution in a magnetic field applied along the magnetic easy axis (i.e. $c$ axis) at LET. In Figure~\ref{spectra_LET} (a)-(d), we show the false-color plots of the energy-momentum cuts of the low-energy magnetic excitation spectra measured at $T$ = 50 mK and in a field of $\mu_0 H$ = 0, 0.25, 0.5, and 0.75 T, respectively.
The spectra were plotted after subtracting the background scattering from the sample environment measured at $\mu_0 H$  = 7 T (note there is no magnetic excitation below 1 meV at 7 T). As before, to enhance the counting statistics, the spectra presented have been averaged over equivalent paths in the reciprocal space as illustrated in Supplementary Figure 4. 
Typical averaged energy-dependent cuts of the spectra at the $\textrm{K}$ and $\textrm{M}$ points are shown in Figure~\ref{spectra_LET_E_cut}. The energy resolution of this dataset ($\Delta E \approx 40$ $\mu$eV) is lower than that obtained at AMATERAS [Figure~\ref{AMA_high_resolution} {and Figure~\ref{AMA_E_cuts}}]. While the low-energy spectrum at zero magnetic field [Figure~\ref{spectra_LET}(a) and Figure~\ref{spectra_LET_E_cut}(a),(e)] appears qualitatively similar in both measurements, the application of a magnetic field drastically changes the excitation spectrum.

In a field of just $0.25 ~\textrm{T}$, the energy of the pseudo-Goldstone excitation branch increases, with a gap of $E = 0.16$ meV at the $\textrm{K}$ point, while the gapless Goldstone mode persists, as shown in Figures~\ref{spectra_LET}(b) and \ref{spectra_LET_E_cut}(b).   
The splitting between these two branches is prominent near the
K point, whereas they appear to remain nearly degenerate at the 
M point.
The roton-like dispersion minimum in the zero-field gapless branch [Figure~\ref{spectra_LET}(a)] is progressively suppressed with increasing magnetic field [Figure~\ref{spectra_LET}(b)], consistent with the system becoming more semi-classical as it approaches the 1/3 magnetization plateau. In contrast, the dip at the 
M point persists for the gapped branch [Figure~\ref{spectra_LET}(b)].
In addition, at zero field there is another dip in the lower boundary of the continuum, located halfway between the 
$\Gamma$ and $\textrm{K}$ points  [see Figure~\ref{spectra_LET}(a) and Figure~\ref{AMA_high_resolution}]. This dip is absent in the linear spin-wave theory calculations shown by solid lines, and is completely suppressed at $\mu_0 H=0.25 ~\textrm{T}$ [Figure~\ref{spectra_LET}(b)].
In the meantime, the excitations become sharper and more intense as they approach the $\Gamma$ point. While LSWT accurately reproduces the slope of the excitations in the $E \rightarrow 0$ limit at both the $\Gamma$ and $\textrm{K}$ points, notable discrepancies arise between the measured and predicted excitations at higher energies.

At $\mu_0 H = 0.5$ T, the magnetic excitation spectrum undergoes further reconstruction, as shown in Figure~\ref{spectra_LET}(c). The gapped excitation at the $\textrm{K}$ point shifts to a higher energy of 0.26 meV [see also Figure~\ref{spectra_LET_E_cut}(c)]. Most remarkably, the continuum of excitations moves towards higher energy with increasing magnetic field, and appears to be associated with the gapped excitation branch. Moreover, the degeneracy at the $\textrm{M}$ point is completely lifted, with the energy of the excitation in the gapped branch increased to 0.35 meV [Figure~\ref{spectra_LET_E_cut}(g)]. The roton-like dip in the gapped branch becomes hardly visible [Figure~\ref{spectra_LET}(c)]. Overall, the excitation intensity becomes much weaker at this field (note the $\times$ 2 factor in color scale).

Finally, the continuum of excitations disappears at $\mu_0 H = 0.75$~T, just below the quantum phase transition to the uud plateau phase at $\mu_0 H_c \approx 0.8$~T, as shown in Figure~\ref{spectra_LET}(d). The excitations evolve into sharp magnons, resembling those observed in the plateau phase at $\mu_0 H_c = 1.5$~T, as previously reported~\cite{Zhu2024}. The gapless branch is well described by LSWT (solid lines). It is worth noting that, LSWT predicts a lower critical field implying that the calculated magnon bands in Figure~\ref{spectra_LET}(d) are gapped. This reaffirms that the continuum of excitations in the supersolid state of \KSeCo is linked to the ``superfluidity'' of the transverse $xy$-moment in the Y-structure.

In contrast, the gapped branch exhibits minor deviations from the LSWT predictions [Figure~\ref{spectra_LET}(d)]. These discrepancies arise because LSWT is not fully accurate near the quantum critical point. Although the spin structure closely resembles the collinear uud configuration, quantum fluctuations significantly reduce the ordered moment from its saturated value \cite{Zhu2024}. Furthermore, the scattering appears diffusive: the energy width of the excitations slightly exceeds the resolution [Figure~\ref{spectra_LET_E_cut}(h)], and additional weak scattering is observed at the $\textrm{K}$ point near 0.5 meV, just below the gapped excitation branch [see Figure~\ref{spectra_LET}(d) and Figure~\ref{spectra_LET_E_cut}(d)].
To further highlight the distinctions of the measured excitation spectra from LSWT, the dynamical spin structure factor (DSSF) calculated using LSWT are shown in Supplementary Figure 6 (left 4 panels).

\subsection{Hamiltonian model\label{sec:theory}}

In the study of real magnetic systems, we encounter two key challenges: extracting a model that accurately describes the most relevant aspects of the system's physics, and then solving it. This task is ``ouroboric'' in nature, as determining the model's parameters requires solving it first and comparing the results with experimental observations. Since most models cannot be solved exactly, this makes the process highly non-trivial. A reasonable approach is to propose a minimal model that captures the essential physical features of the system, at least within a certain energy scale, and solve it using well-justified approximations. Following this methodology, and building on the results outlined in the previous sections and in our previous work \cite{Zhu2024}, we propose modeling the magnetic properties of \KSeCo using an XXZ Hamiltonian,
\begin{equation}    
\mathcal{H} = \sum_{\langle i,j\rangle } \left[J_{zz} S^z_iS^z_j + J_{xy}(S^x_iS^x_j+ S^y_iS^y_j)\right] -  h \sum_i S^z_i,
\label{Eq:Ham}
\end{equation}
where $h = g_{zz} \mu_B \mu_0  H$, $\mu_0 H$  is the external magnetic field, $\mu_B$ is the Bohr magneton and $g_{zz} = 7.9$ the ${z}$-component of the Landé g-tensor. The exchange interactions determined in \cite{Zhu2024} through high-field magnetometry measurements were further refined by detailed comparison with theoretical calculations, as discussed below. This process yielded updated values of \( J_{zz} = 3.1 \, \text{meV} \), \( J_{xy} = 0.217 \, \text{meV} \), and an anisotropy ratio of \( \alpha = J_{xy}/J_{zz} = 0.07 \). The refinement was essential to achieve a precise agreement with the specific heat and magnetization curves presented in subsequent sections.

We begin by examining the zero-temperature classical and semi-classical predictions for the Hamiltonian model in Eq.~\ref{Eq:Ham} and assessing their agreement with experimental observations. Given the small ratio \(\alpha\), it is natural to first consider the strictly Ising limit (\(J_{xy} = 0\)). In the absence of an external magnetic field (\(H = 0\)), the Hamiltonian reduces to an antiferromagnetic triangular Ising model, which exhibits an extensively degenerate ground-state manifold \cite{wannier1950}. This manifold is spanned by configurations in which no triangular plaquette has all three spins aligned in parallel. As introduced earlier, we refer to this subspace as the ``Wannier space'' $\mathcal{W}$.

An infinitesimal magnetic field applied along the $z$-direction lifts this degeneracy, stabilizing three distinct ``uud'' ground states, where two spin sublattices align parallel to the field while the third sublattice orients antiparallel to it. Consequently, the magnetization undergoes a discontinuous jump from zero to one-third of the saturation value.

As we move away from the Ising limit (\(J_{xy} \ll J_{zz}\)) and in the absence of an external field, the ground-state manifold of the classical spin (\(S \to \infty\)) Hamiltonian becomes two-dimensional. The first dimension emerges from the continuous U(1) Hamiltonian symmetry associated with global spin rotations around the $z$-axis. The second dimension arises from an accidental continuous degeneracy, which corresponds to the freedom in selecting one of the three polar angles for the spins within the magnetic unit cell \cite{Kleine1992, Sheng1992, Kawamura1985, Murthy1997}. These angles, measured relative to the \(z\)-axis, as shown in Figure~\ref{Fig:Y}(b), satisfy the following relationships:

 \begin{align}
     \vartheta_B &= \varepsilon + \delta, \ \ \ \vartheta_C = \varepsilon - \delta \nonumber \\ 
     \cos{\delta} &= \frac{\alpha}{(1+ \alpha) [\sin^2{\vartheta_A} + \alpha\cos^2{\vartheta_A} ]^{1/2} } \nonumber \\
     \cos{\varepsilon} &= \frac{\alpha \cos{\vartheta_A}}{[\sin^2{\vartheta_A} + \alpha\cos^2{\vartheta_A} ]^{1/2}} 
     \label{Eq:ang_rel}
 \end{align}
 with $\vartheta_A$ arbitrarily chosen. Eventually,  the application of an infinitesimal magnetic field along the $z$-direction lifts the accidental degeneracy, selecting the Y-shaped ground state configuration depicted in Figure~\ref{Fig:Y}(a).

\begin{figure}[t!]
    \centering
     \includegraphics[width=\columnwidth]{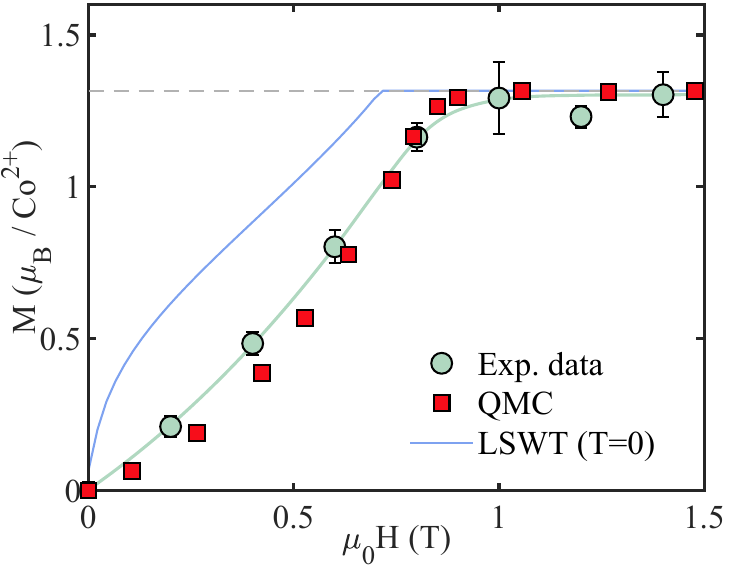}
    \caption{\textbf{Field dependence of magnetization in \KSeCo.} Magnetization of $\text{K}_2\text{Co}(\text{Se}\text{O}_3)_2$ as a function of applied field at $T=0.19~\text{K}$. Circles are experimental data from Ref.~\cite{Zhu2024}. The solid curve is a guide for the eye. Red squares are QMC results (this work). The blue curve is the LSWT calculation. The dashed line represents one-third of the experimental saturation magnetization.}
    \label{Fig:magcurve}
\end{figure}

Although the classical ground state of the Hamiltonian in Eq.~\ref{Eq:Ham} qualitatively matches the state of \KSeCo, it completely fails to predict the magnetization at zero field. The classical solution gives a magnetization of
\begin{equation}
    \frac{M(H=0)}{M_{\rm sat}} = \frac{1}{3}~ \frac{1-\alpha}{1+\alpha} \  \xrightarrow{\alpha \ll 1}  \frac{1}{3}
\end{equation}
which stands in stark contrast to the zero magnetization observed experimentally, as shown in Figure~\ref{Fig:magcurve}. 
Although we cannot entirely rule out the possibility of antiferromagnetic interlayer correlations characterized by a coupling constant $J_{\perp}$, the absence of a jump or noticeable change in the slope of the magnetization curve suggests that each individual layer has zero net magnetization.

Moving now to a semi-classical approach, spin-wave calculations by Kleine et al.~\cite{Kleine1992} demonstrated that quantum fluctuations lift the classical accidental ground-state degeneracy, stabilizing the coplanar Y-order. Additionally, they observed a significant reduction in the total ground-state magnetization, nearly driving it to zero, which aligns closely with our experimental results. However, while the zero-field magnetization predicted by LSWT is close to our experimental findings, the magnetization curve $M(\mu_0H)$ exhibits a significant discrepancy with the experimental data [see Figure~\ref{Fig:magcurve}], particularly in predicting a lower critical field for the transition to the 1/3 plateau.

At low energies (\(\omega \sim J_{xy}\)) and zero field, spin-wave calculations predict two degenerate magnon bands. One corresponds to a Goldstone mode associated with the broken U(1) symmetry, arising from the selection of a specific orientation for the spins in the \(xy\)-plane. The other is a “pseudo-Goldstone” mode, resulting from the accidental continuous degeneracy of the classical ground state [see solid lines in Figure~\ref{AMA_high_resolution}]. Higher-order $1/S$ corrections are expected to gap out the pseudo-Goldstone mode~\cite{Savary2012,Rau2018}. This is in qualitative agreement with our ultra-high-resolution inelastic neutron scattering measurement, where a tiny energy gap of 60 $\mu$eV is detected at the K point [Figure~\ref{AMA_high_resolution} inset]. At higher energies (\(\omega \sim J_{zz}\)), a third, nearly dispersionless band emerges [Figure~\ref{AMA_3meV}]. This mode corresponds to excitations into states within the subspace \(\mathcal{W}^{\perp}\), which is orthogonal to the Wannier subspace \(\mathcal{W}\), as illustrated in Figure~\ref{Fig:High_ene_exitations}.

While LSWT accurately describes the excitation velocity near the K-point, it fails to capture the local minima in the dispersion around the M-point observed in the INS data [Figure~\ref{AMA_high_resolution}]. Moreover, the significant reduction in the ordered moment found by Kleine suggests that a $1/S$ expansion is inadequate for accurately describing the dispersion and spectral weight distribution of shorter-wavelength excitations. In particular, the first quantum correction to the dipole moment renormalization is nearly as large as the full moment, underscoring the absence of a suitable expansion parameter in this approach.

It is now insightful to examine the thermodynamics, i.e., the \(T > 0\) regime. As calculated by Wannier, the pure Ising model at zero field exhibits a macroscopic residual entropy (zero-temperature entropy) of \(\Delta S/R \approx 0.323\), approximately half of \(\ln 2\)—the full saturation entropy of a spin-\(\frac{1}{2}\) system.  

In our model, quantum fluctuations arising from first-order corrections in \(J_{xy}\) are expected to lift the extensive ground-state degeneracy by selecting a specific ground-state submanifold within the Wannier subspace. Consequently, the Wannier entropy should be recovered at a crossover temperature \(T_{\rm cr} \sim J_{xy}/k_B\), where the entropy curve
\begin{equation}
    \Delta S(T) = \int_0^{T} \frac{C_v}{T'} dT'    
\end{equation}
should display a plateau. This behavior is precisely observed in \KSeCo, as shown in Figure~\ref{Fig:Ent_exp}. The relatively narrow temperature range of the plateau can be attributed to the moderate separation of energy scales, with \(J_{xy}\) being ``only'' an order of magnitude smaller than \(J_{zz}\).

As demonstrated in the seminal work of José \textit{et al.}~\cite{Jose77}, identifying the different spontaneously broken symmetries is essential for understanding phase transitions in spin models. Given the distinct energy scales governing Ising (solid) and XY (BEC) orderings, the former occurs at a higher critical temperature.  

\begin{figure}[t!]
    \centering
     \includegraphics[width=\columnwidth]{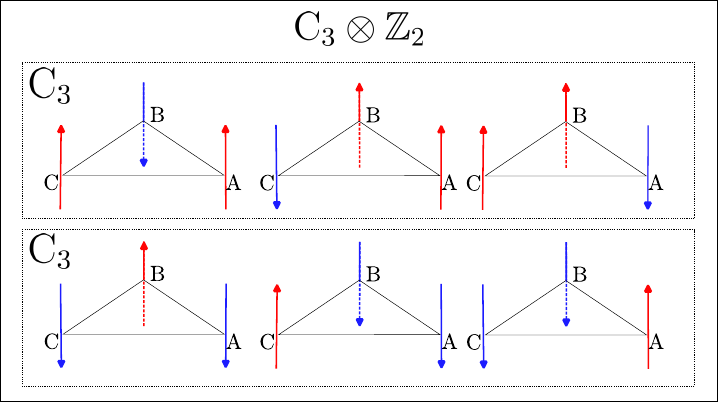}
    \caption{\textbf{Discrete lattice symmetries.} Sketch of the discrete symmetries broken during the emergence of Ising (solid) order at zero field (large box) and at non-zero field (small boxes). The choice between the top or bottom small box depends on the orientation of the external field.}
    \label{Fig:disc_sym}
\end{figure}

At zero field, the Ising ordering breaks the discrete \(\text{C}_3\) lattice symmetry (associated with \(2\pi/3\) rotations) by selecting one of the three sublattices to host the minority spin. It also breaks time-reversal (\(\mathbb{Z}_2\)) symmetry, as the minority spin can point either up or down [see Figure~\ref{Fig:disc_sym}]. Consequently, this ordering emerges through two Berezinskii-Kosterlitz-Thouless (BKT) transitions at \(T_{\text{BKT},1}\) and \(T_{\text{BKT},2}\), with \(T_{\text{BKT},1} > T_{\text{BKT},2}\). A critical phase exists for \(T_{\text{BKT},2} < T < T_{\text{BKT},1}\), while below \(T_{\text{BKT},2}\), a massive phase with long-range Ising order sets in. This behavior is evident in our high-temperature specific heat data [Figure~\ref{Fig:high_T_cv}(b)].  

At even lower temperatures, the XY order—associated with U(1) spin symmetry breaking—undergoes a separate BKT transition at \(T_{\text{BKT},3} \ll T_{\text{BKT},2}\). In our experimental data, this transition appears as a subtle elbow in the inset of Figure~\ref{Fig:high_T_cv}(b).  

Since only the \(\text{C}_3\) symmetry breaks for finite magnetic field, the critical phase acquires long-range Ising order and merges with the massive phase below \(T_{\text{BKT},2}\). Simultaneously, the two BKT transitions are replaced by a single second-order phase transition in the 2D Potts universality class at \(T_{\text{c}1}\), which converges to \(T_{\text{BKT},1}\) as the field vanishes.

Although real systems, which are inherently three-dimensional, are expected to exhibit a single phase transition at \(T_{c1} \simeq T_{{\rm BKT},1}\) replacing the upper two transitions, and a three-dimensional XY phase transition at \(T_{c2} \simeq T_{{\rm BKT},3}\) replacing the lower BKT transition, the anomaly in the specific heat associated with these transitions is too subtle to be detected in our experiments. Consequently, \KSeCo approximates extremely well the expected phase transition of the 2D XXZ model (Eq.~\ref{Eq:Ham}) and shows qualitative consistency with the predictions of previous studies~\cite{Kawamura1985, Stephan2000, Sheng1992, Melchy2009}.

Up to this point, we have presented arguments, supported by both the current literature and our experimental data, demonstrating that \KSeCo is well described by the XXZ Hamiltonian (Eq.~\ref{Eq:Ham}). Specifically, we highlighted that LSWT accurately captures the renormalization of the ordered moment at zero temperature and zero field but fails to reproduce the \(M(\mu_0 H)\) curve. Additionally, while thermal fluctuations are qualitatively well described by a classical spin model, our INS data reveal that quantum fluctuations significantly impact the excitation spectrum, causing deviations from the semi-classical prediction. This suggests the need for a fully quantum approach to properly capture the system's behavior. Although a \(1/S\) expansion could be pursued, its convergence is uncertain due to the lack of a well-defined small expansion parameter. In the remainder of this work, we will take a different approach to solving the full quantum Hamiltonian.

\subsection{Low energy theory}
Exploiting the clear energy separation discussed earlier, explicitly reflected in the ratio $\alpha$, we treat this parameter as a perturbative one. This approach follows the initial observations of A. Sen \textit{et al.}~\cite{Sen2008} for the non-frustrated case ($J_{xy}<0$), later independently generalized by Fa Wang\cite{Wang2009} and H. C. Jiang~\cite{Jiang2009} to the frustrated regime.

To capture the low-energy physics, we construct an effective Hamiltonian, ${\cal H}_{\text{eff}}$, by projecting the original Hamiltonian (Eq.~\eqref{Eq:Ham}) onto the Wannier subspace ${\cal W}$. Wang \textit{et al.} demonstrate the existence of a unitary transformation that reverses the sign of the $J_{xy}$ term in ${\cal H}_{\text{eff}}$ while preserving both the eigenvalue spectrum and diagonal correlations in the $S^z$ basis. This transformation eliminates the QMC sign problem, allowing us to harness this powerful numerical technique to investigate both static and dynamic properties. In the following, we provide a detailed discussion of this transformation.

\begin{figure}[t!]
    \centering
    \includegraphics*[width=0.99\columnwidth]{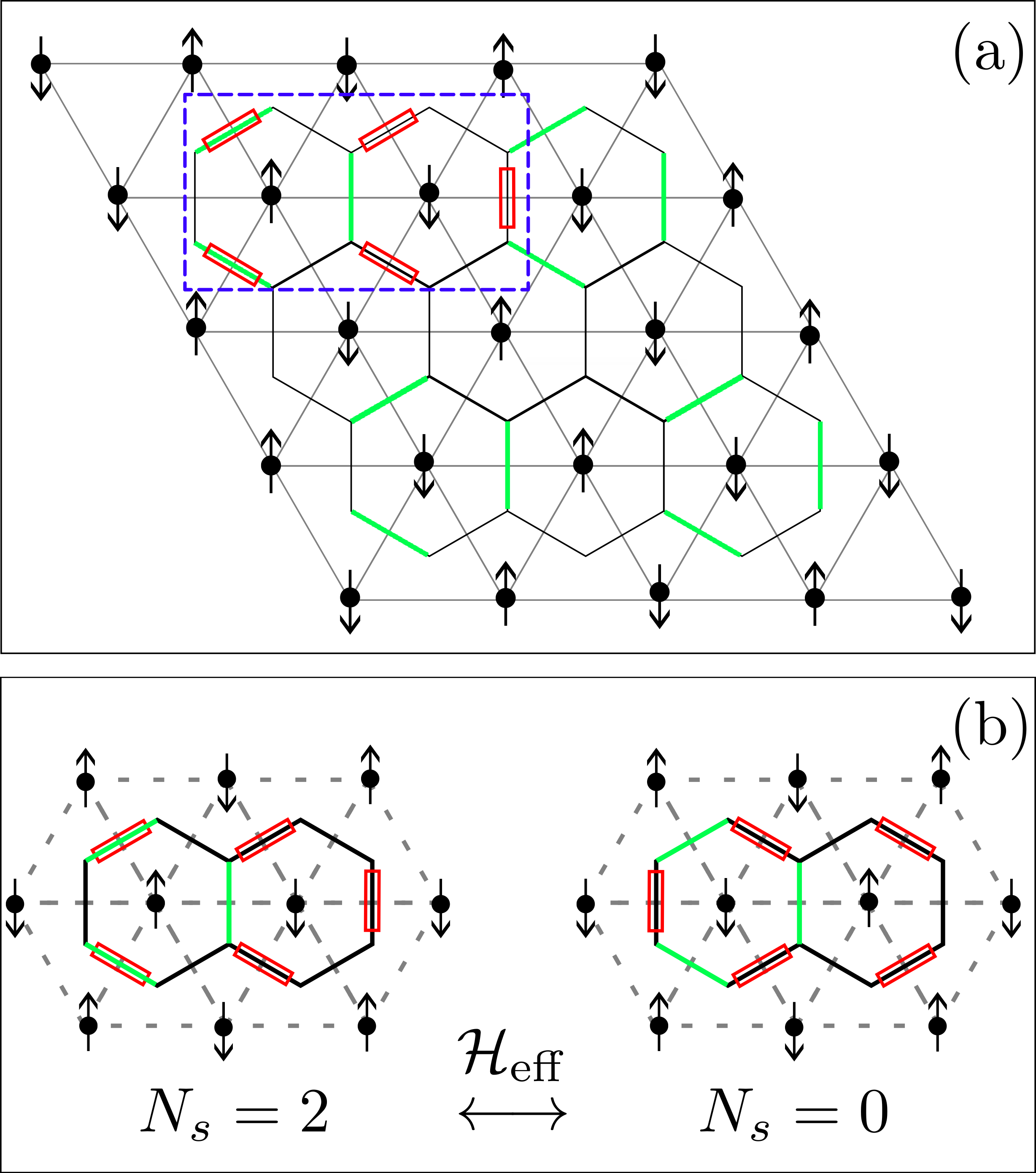}
    \caption{\textbf{Dual lattice and dimer model.} a) Triangular lattice (grey line) and the associated honeycomb dual lattice (bold line). The green bonds denote the ``special'' ones used in the unitary transformation. The blue dashed lines show a resonant hexagon and the red boxes represent the dimer covering. b) Effect of the effective Hamiltonian over a resonant hexagon, the same that in Figure (a),  the number of special bonds covered by dimers is represented by $N_s$.}
\label{Fig:dual}
\end{figure}

Since the Hilbert space of the effective low-energy model is the Wannier subspace ${\cal W}$, it is first necessary to introduce a suitable algebra of operators that act on this highly constrained space. This is achieved by introducing dimer coverings on the dual honeycomb lattice. As shown in Figure~\ref{Fig:dual}(a), a dimer is placed on the dual link perpendicular to each frustrated bond (a bond connecting two parallel spins) of a given spin configuration. We denote a specific dimer covering by ${\cal C}$. Because we are focusing on only one of the two states related by time-reversal symmetry, the proposed duality transformation results in a 2:1 mapping.

 In this language, ${\cal H}_{\rm eff}$ is simply a quantum dimer model with a ring-exchange term that acts on each pair of adjacent hexagons on the dual honeycomb lattice (corresponding to each bond $\langle i,j \rangle$ of the triangular one).

\begin{equation}
    {\cal H}_{\rm eff} = \frac{J_{xy}}{2} \sum_{\doublehex_{\langle ij\rangle}} (|\maxflip_{\langle ij \rangle} \rangle \langle \maxfliptwo_{\langle ij \rangle} | + h.c.)
\label{Eq:dimer}
\end{equation}

In other words, ${\cal H}_{\rm eff}$ connects basis states that present a double-hexagon resonance as sketched in Figure~\ref{Fig:dual}(b). The ground state is then obtained by optimizing the energy gain produced by this dimer resonance. 

The most general eigenstate of the effective Hamiltonian is written as a linear superposition of the different dimer coverings,
\begin{equation}
    \ket{\Psi} = \sum_{\mu} \psi_{\mu} \ket{\cal C_{\mu}}
\end{equation}
with $\ip{{\cal C}_{\nu}}{{\cal C}_{\mu}} = \delta_{\nu \mu}$ and $\sum_{\mu} |\psi_{\mu}|^2=1$. If a unitary transformation \(\hat{U}\) exists that reverses the sign of \({\cal H}_{\rm eff}\), the low-energy model can be mapped onto one with a negative \(J_{xy}\), corresponding to a non-frustrated XY interaction. Consider the lattice depicted in Figure~\ref{Fig:dual}(a), where specific bonds—designated as ``special''—are marked with green lines. For any given dimer covering $\cal C$, we define $N_{s}({\cal C})$ as the number of special bonds occupied by dimers. As illustrated in Figure~\ref{Fig:dual}(b), a resonance within any resonant hexagon changes $N_{s}({\cal C})$ by $\pm 2$. To capture this behavior, we introduce the following unitary transformation on the dimer basis~\cite{Wang2009}:
\begin{equation}
\ket{ \mathcal{C}^{\prime}} := \hat{U} \ket{\mathcal{C}} = e^{i\frac{\pi}{2}{N_s(\mathcal{C})}}~ \ket{\mathcal{C}}
\end{equation}
where $i$ is the imaginary unit. Consequently, basis elements connected by ${\cal H}_{\rm eff}$ differ in their $N_s$ values by integer multiples of $\pm 2$. Defining $\ket{\Psi'} =  \hat{U} \ket{\Psi}$ and $\ket{\Phi'} =  \hat{U} \ket{\Phi}$, we obtain
\begin{equation}
\begin{split}
\langle {\Phi }'| \mathcal{H}_{\mathrm{eff}}| {\Psi }'\rangle 
&= \sum\limits_{\mu \nu} e^{i\frac{\pi}{2} \left[ N_s(\mathcal{C}_\mu) - N_s(\mathcal{C}_\nu) \right]} \, \phi_\nu^* \psi_\mu \langle \mathcal{C}_\nu | \mathcal{H}_{\mathrm{eff}} | \mathcal{C}_\mu \rangle \\[6pt]
&= \sum\limits_{\mu \nu} e^{\pm i\pi} \, \phi_\nu^* \psi_\mu \langle \mathcal{C}_\nu | \mathcal{H}_{\mathrm{eff}} | \mathcal{C}_\mu \rangle \\[6pt]
&= \sum\limits_{\mu \nu} - \phi_\nu^* \psi_\mu \langle \mathcal{C}_\nu | \mathcal{H}_{\mathrm{eff}} | \mathcal{C}_\mu \rangle \\[6pt]
&= \langle \Phi | -\mathcal{H}_{\mathrm{eff}} | \Psi \rangle
\end{split}
\end{equation}

implying that $\hat{U}$ maps ${\cal H}_{\rm eff}$ into $-{\cal H}_{\rm eff}$. Since the unitary transformation preserves the energy spectrum, thermodynamic properties, which depend solely on the energy eigenvalues, remain unchanged. The same holds for observables that are diagonal in the dimer basis, implying that \(\langle \Psi' | S^z_i(t) S^z_j(0) | \Psi' \rangle = \langle \Psi | S^z_i(t) S^z_j(0) | \Psi \rangle\). However, this does not apply to correlators like \(\langle S^x_i(t) S^x_j(0) \rangle\) or \(\langle S^y_i(t) S^y_j(0) \rangle\), which take a more complex form after the transformation.

\begin{figure}[t!]
    \centering
    \includegraphics[width=0.7\columnwidth]{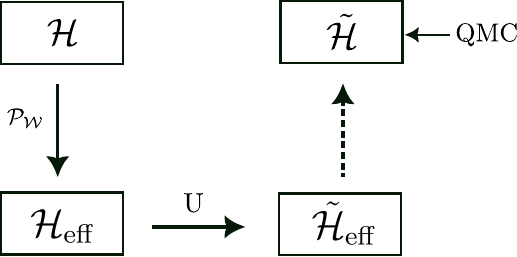}
    \caption{\textbf{Sketch of the strategy used to attack the present problem.} ${\cal P}_{\cal W}$ represent the projection of the full Hamiltonian over the Wannier space to arrive to a low energy Hamiltonian ${\cal H}_{\rm eff}$, that is mapped by the unitary transformation $U$ to an non-frustrated Hamiltonian ${\Tilde{\cal H}}_{\rm eff}$. We promote ${\Tilde{\cal H}}_{\rm eff}$ to ${\Tilde{\cal H}}$ in the complete Hilbert space and solve it using QMC.  }
    \label{Fig:Ham_mapping}
\end{figure}

We summarize our strategy for solving the Hamiltonian ${\cal H}$ in Figure~\ref{Fig:Ham_mapping}. To address the sign problem, the original XXZ Hamiltonian is projected onto the Wannier space where the transformation introduced by Wang permits reverse the sign of the $xy$-term without altering the thermodynamic properties or the $\langle S_{i}^z(t)S_{j}^z(0)\rangle$ correlation functions. While implementing an efficient QMC algorithm under the constraints of ${\cal W}$ implied a significant computational challenge, we bypass this difficulty by solving the non-frustrated XXZ Hamiltonian and neglecting second-order corrections in $J_{xy}$. The validity of this approximation is demonstrated by the results discussed in the following sections.

\subsection{QMC simulations}

To further validate the proposed spin Hamiltonian and the low-energy model derived in the previous section, we computed the uniform magnetization curve $M(\mu_0 H)$, up to the values corresponding to the onset of the $1/3$ magnetization plateau.  As shown in Figure~\ref{Fig:magcurve}, the calculated magnetization curve is in good agreement with the experimental data, predicting a critical field $\mu_0 H_{c} \approx 0.85~\textrm{T}$, which matches well with the observed value. The theoretical calculation shows a slightly non-linear increase of $M$ with $H$, which is characteristic of 2D systems but seems to be absent in the measured curve. Small discrepancies with the experimental data are likely due to the fact that the current mapping is exact only in the Ising limit and at zero field. As a result, small higher-order corrections in $J_{xy}$ are expected to introduce discrepancies with the experimental results. While we do not expect a perfect match, we anticipate a good agreement, given the small ratio \(\alpha\) and the low field, which makes it plausible that the main contributions come from the Wannier states.

\begin{figure}[t!]
    \centering
     \includegraphics[width=1\columnwidth]{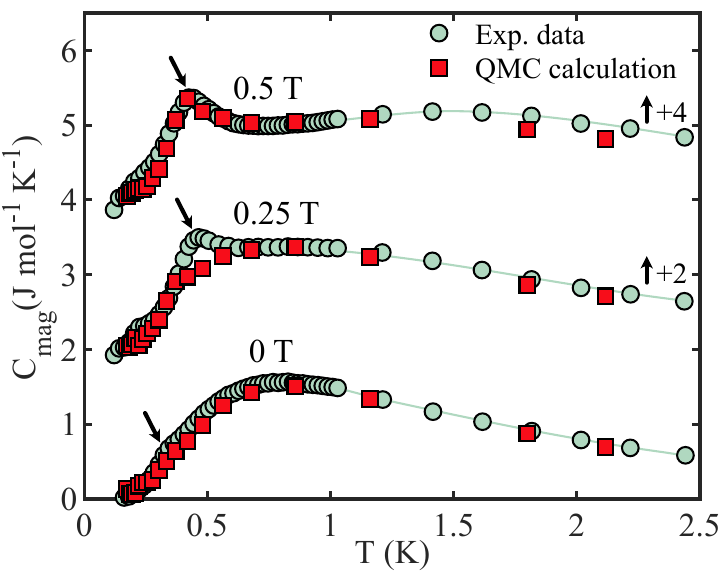}
    \caption{\textbf{Experimental and calculated magnetic specific heat of \KSeCo.} Magnetic specific heat of $\text{K}_2\text{Co}(\text{Se}\text{O}_3)_2$ for different fields $\mu_0 H$ = 0, 0.25 and 0.5~T applied along the $c$ axis. Circles: experimental data from Ref.~\cite{Zhu2024} with phonon and nuclear contribution subtracted as described in the text. The line is a guide for the eye. Squares: QMC calculations (this work). The black arrows indicate the BEC transition.}
    \label{Fig:spe_comp}
\end{figure}

The specific heat \( C_v = R~ \partial_T \langle \mathcal{H} \rangle \) is computed for $\mu_0 H=0, 0.25 $ and $0.5$ T and the low temperature interval $0.16~\text{K} < T < 3.27~\text{K}$.  
The comparison with the experiment is presented in Figure~\ref{Fig:spe_comp}. The phonon and Schottky contributions were subtracted from the measured data as in Figure~\ref{Fig:high_T_cv}(b). The mentioned low-temperature broad peaks are successfully reproduced by the QMC simulations of the low-energy effective model, confirming the validity of the  Hamiltonian parameters of \KSeCo. The lowest temperature reached by QMC calculations is slightly higher than the experimental value ($T=0.12$~K). This limitation, combined with finite-size effects, likely explains the minor discrepancies observed in comparisons with experimental data.

\begin{figure}[t]
    \centering
    \includegraphics[width=1\columnwidth]{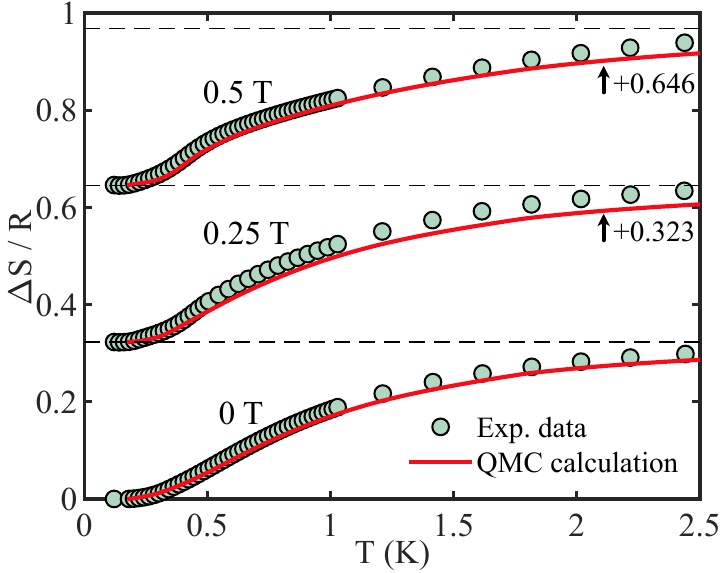}
    
    \caption{\textbf{Experimental and calculated magnetic entropy of \KSeCo. } Temperature dependence of magnetic entropy at $\mu_0 H=0, 0.25$ and $0.5 ~\textrm{T}$. Circles: experimental data obtained by integrating the measured magnetic heat capacity $C_{\text{mag}}/T$. Red lines:  QMC calculation. The dashed lines indicate the Wannier entropy.}
    \label{Fig:entropy_curve}
\end{figure}

Moreover, as shown in Figure~\ref{Fig:entropy_curve}, the low-energy model also reproduces the temperature dependence of the magnetic entropy as it approaches the Wannier value. These results are also in good correspondence with previous Lanczos ones ~\cite{Ulaga2023}.

The low-energy excited states, with energies on the order of $E \sim J_{xy}$, reside within the Wannier space, and it is anticipated that our approach can effectively capture these excitations as well. However, a significant limitation of our methodology is our inability to compute the ${\cal S}^{xx}$ and ${\cal S}^{yy}$ correlation functions. Consequently, we focus on the ${\cal S}^{zz}$ component of the DSSF, defined as
\begin{equation}
    {\cal S}^{zz}({\bf q}, \omega) = \int_0^{\infty} e^{-i\omega t}\langle S_{-\bf q}^z(t)S_{\bf q}^z(0) \rangle ~dt,
    \label{Eq:dssf}
\end{equation}
where $S^z_{{\bf q}} = \frac{1}{\sqrt{{\cal N}}}\sum_j e^{-{\bf q}\cdot {\bf r}_j}S^z_j $, ${\cal N}$ is the number of sites and  $j$ runs over all lattice sites.

Although a direct comparison with experimental data is not possible—since it includes all three components of the dynamical spin structure factor—the strong anisotropy of the \( g \)-tensor, discussed in Section~\ref{Sec:ins}, indicated that the calculation should account for most of the measured spectral weight. Several experimentally observed features are clearly reflected in our simulations, highlighting the strong quantum nature of the magnetic excitations. 

\begin{figure}[!t]
    \centering
        \includegraphics[width=0.98\columnwidth]{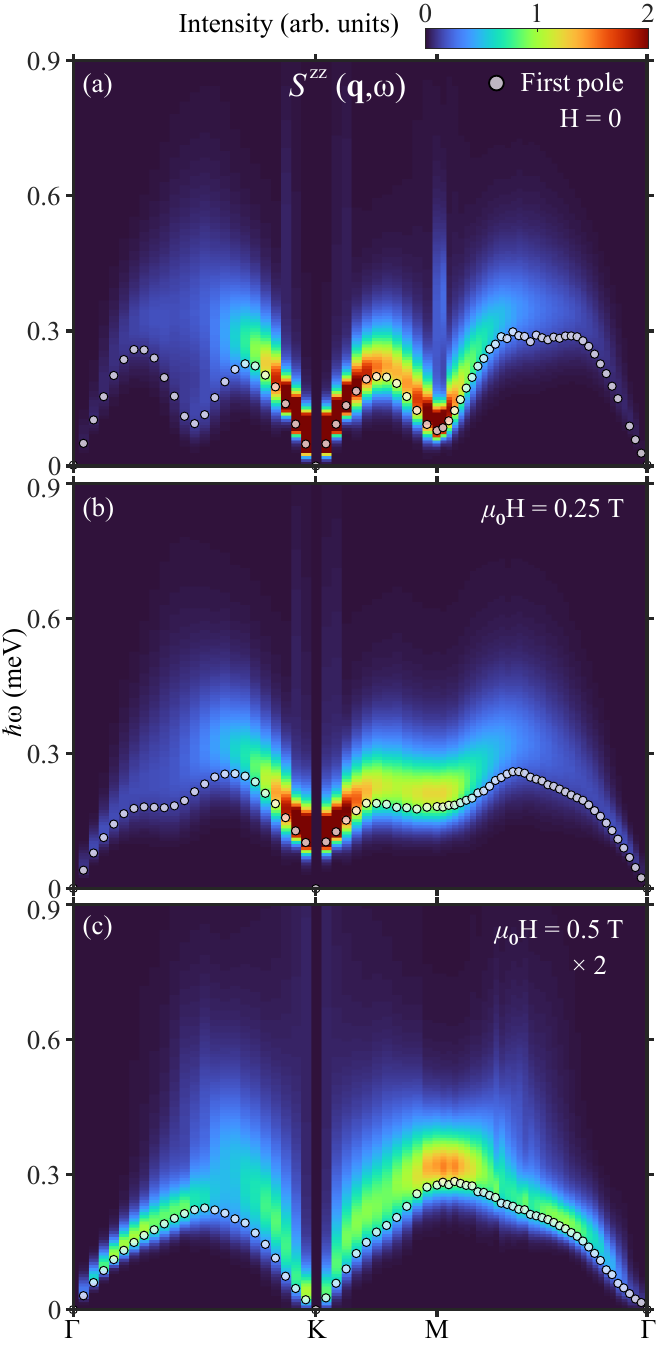}
    \caption{\textbf{Evolution of the ${\cal S}^{zz}$ component of the DSSF in the supersolid phase in a magnetic field, obtained by QMC calculations.} ${\cal S}^{zz}$ component of the dynamical spin structure factor computed with QMC and analytic continuation for (a) $\mu_0 H=0$ T, (b) $\mu_0 H=0.25$ T and (c) $\mu_0 H=0.5$ T. The white dots are the first poles of ${\cal S}^{zz}$ extracted with the method presented in Supplementary Note 5.}
    \label{Fig:QMC_spectrums}
\end{figure}

Figure~\ref{Fig:QMC_spectrums} presents QMC simulations for \(\mu_0 H = 0\), \(0.25\), and \(0.5~\textrm{T}\). The white dots indicate the lowest energy pole (first pole) of the DSSF extracted from the QMC data as presented in Methods. Since the QMC simulations are performed in imaginary time, an analytic continuation is applied to transform the dynamical correlation function to the frequency domain. However, we encountered convergence issues at the Goldstone modes, leading us to exclude this data from the figures. Nevertheless, the existence of a zero mode is confirmed through the extraction of the first pole. Further details on the calculations are provided in Methods.

The key features of the experimental spectrum are well reproduced at zero field [Figure~\ref{Fig:QMC_spectrums}(a)]. Our simulations accurately capture the roton-like minima, and the continuum along the \(\Gamma\)–K path shows strong agreement with the experimental data. In Figure~\ref{AMA_E_cuts}, we plot the calculated momentum-resolved cuts (red solid lines), highlighting the precise prediction of both the potential quasiparticle peak and the broad continuum. To ensure consistency, intensities are rescaled to match the \(\textrm{M}\) point. Additionally, the bottom edge of the continuum is well described by the first pole of the QMC data [Figure~\ref{AMA_high_resolution}, \ref{spectra_LET}(a)], with a perfect match in the excitation velocity near K. A significant discrepancy with the experimental data is the absence of an extended continuum at the $\textrm{M}$ point in the simulations. We conjecture that this continuum primarily originates from the ${\cal S}^{xx}({\bf q}, \omega)$ and ${\cal S}^{yy}({\bf q}, \omega)$ components of the dynamical spin structure factor, a hypothesis supported by the Lanczos results reported in \cite{Ulaga2024}.

The comparison becomes more challenging at finite fields, because the magnetic structure continuously evolves into the collinear up-up-down phase, implying that most of the spectral weight of ${\cal S}^{zz}({\bf q}, \omega)$ comes from longitudinal excitations corresponding to the two-magnon continuum, while the spectral weight from transverse modes is mainly captured by the ${\cal S}^{xx}({\bf q}, \omega)$ and ${\cal S}^{yy}({\bf q}, \omega)$ components. 
While this creates a noticeable discrepancy, we can still observe in Figure \ref{Fig:QMC_spectrums}(b) a clear energy splitting of the pseudo-Goldstone mode and the flattening of the roton minimum at $\mu_0 H = 0.25 ~\textrm{T}$. Finally, the lower band is partially recovered at $\mu_0 H = 0.5~\textrm{T}$ [Figure \ref{Fig:QMC_spectrums}(c)].

Figure~\ref{spectra_LET_E_cut}(e-g) compares the QMC calculations with experimental data at the M-point, demonstrating that our simulations accurately predict the location of the lowest-energy quasiparticle peak. This agreement is further supported by the good match between the first pole and the bottom of the continuum in Figure~\ref{spectra_LET}.

To highlight the differences with LSWT, Supplementary Figure 6 shows the total DDSF and its \({\cal S}^{zz}\) component calculated using this approach.

\section{Discussion}

 Throughout this work, we presented strong arguments supporting that the XXZ Hamiltonian (Eq.~\eqref{Eq:Ham}) and its corresponding low-energy theory, valid in the small \( J_{xy}/J_{zz} \) limit, accurately describe the low-temperature magnetic properties of \KSeCo. While we cannot completely rule out additional interactions, the good agreement between theory and experiment suggests that such terms are likely very weak and would only become relevant at extremely low temperatures or energy scales. Therefore, at studied energy scales, we can conclude with reasonable confidence that \KSeCo exhibits a spin-supersolid phase, which can be described by a nearest-neighbor XXZ Hamiltonian with strong easy-axis anisotropy. 
 We emphasize the necessity of moving beyond semiclassical approaches to accurately capture the excitation spectrum of this model. While our analysis does not explicitly determine the nature of the continuum, it clearly demonstrates that it emerges from strong quantum fluctuations.

Even more remarkable is the prospect of identifying a material that serves as an ideal platform for exploring this physics. \KSeCo\ not only enables high-resolution experimental investigations of the excitation spectrum using modern neutral spectroscopy techniques but also provides a rare opportunity to apply QMC methods, which are typically inaccessible for general frustrated magnetic systems.
 
 Future work can now address new questions, such as: What is the nature of the low-energy excitation continuum? Can it be captured by a renormalized semiclassical theory incorporating higher-order $1/S$ corrections \cite{Chernyshev2009}, or does it originate from weakly confined spinon excitations, better described by large-$N$ approaches \cite{Ghioldi2018, Bose2025}?
 
 A few theoretical studies \cite{jia2023, Gao2024, Chi2024} have recently investigated the structurally similar triangular lattice antiferromagnet \NaBaCoP, which is described by the same easy-axis XXZ Hamiltonian \cite{Gao2022}. The ground state of \NaBaCoP has also been proposed to be a spin supersolid with a ``Y" structure \cite{Gao2022, Xiang2024}, and inelastic neutron scattering measurements reveal a broad continuum of magnetic excitations \cite{sheng2024,Gao2024}. The key distinction lies in the exchange anisotropy ratio: \(\text{\NaBaCoP}\) has \(\alpha \approx 0.6\), significantly further from the Ising limit than \(\text{\KSeCo}\) (\(\alpha \approx 0.07\)). Consequently, our theoretical approach is not suitable for \(\text{\NaBaCoP}\) due to the lack of a well-defined expansion parameter. Moreover, the exchange interactions are much weaker, with \(J_{zz} = 0.125 \, \text{meV}\) \cite{Sheng2022, sheng2024}, which results in a lower characteristic energy scale for the magnetic exciations, making it even more challenging to resolve the structure of the excitations by neutron spectroscopy.

Nevertheless, we may still gain some insights on the nature of the scattering continuum in \KSeCo by comparing our measured spectra with the theoretically predicted ones qualitatively. One proposed scenario suggests that the spin supersolid ground state is a precursor to a Dirac spin liquid \cite{jia2023}. In this framework, \black{a Dirac spin liquid state coexists with a supersolid order.} The continuum of excitations arises from fractionalized spinons in the spin liquid state. The calculated excitation spectrum is qualitatively similar to the experimental observations: a continuum of excitations bounded by linear dispersion at the $\textrm{K}$ and $\Gamma$ points. The calculated constant-energy slice of the magnetic spectrum shows excitation pockets centered not only at the $\Gamma$ and $\textrm{K}$ points, but also near the $\textrm{M}$ point and the wave vector halfway between $\Gamma$ and $\textrm{K}$ \cite{Zhu2024}. \black{However, the excitation spectrum of the precursory Dirac spin liquid is predicted to develop a small gap due to the presence of the supersolid order. While at the $\textrm{M}$ point, we experimentally observe a gapped, roton-like dip consistent with the findings in \cite{jia2023}, the measured spectrum remains gapless at the $K$ and $\Gamma$ points.}

Other theoretical works offer alternative interpretations. \black{A recent DMRG study shows that the spin supersolid phase in easy-axis triangular lattice antiferromagnets exhibit a double magnon-roton dispersion at low energy $-$ one branch with true Goldstone mode and the other with a gapped pseudo-Goldstone mode. Roton-like dispersion dips are present in both branches at the $M$ point and are nearly degenerate. Besides these spin-wave excitations, DMRG further predicted excitation continuum which remains even when pushing to rather high energy resolution in numeric simulations \cite{Gao2024}. The peculiar double magnon-roton structure in the low-energy excitations is also obtained in a tensor-network renormalization study \cite{Chi2024}. }
The gapless branch is attributed to fluctuations perpendicular to the plane of the ordered moment (purely transverse), while the gapped one is associated with fluctuations in the plane of the Y-structure. 
This agrees well with our observations in \KSeCo [see Figure~\ref{AMA_high_resolution} and inset]. In \NaBaCoP, the pseudo-Goldstone mode is predicted to be around 0.01 meV \cite{Chi2024}, which is extremely difficult to identify experimentally. In contrast, in \KSeCo, thanks to the high energy resolution of the AMATERAS spectrometer, the pseudo-Goldstone mode as low as 0.06 meV can be resolved [Figure~\ref{AMA_E_cuts}(a)].

\black{Most remarkably, with improved energy resolution, the tensor network study \cite{Chi2024} found that the magnetic excitation spectrum of the spin supersolid displays discrete magnon modes that are very close in energy. The observed continuum of excitations \cite{sheng2024} is then ascribed to the broadening due to finite experimental energy resolution. This stands in stark contrast to the spinon picture, where the excitation continuum emerges from fractionalized spin excitations, and points to a completely different scenario for the nature of the excitations \cite{Mingfang2023}. 

In addition, the evolution of the excitation spectrum in a magnetic field has been examined within the same framework.}
It has been predicted that the roton-like dip in the gapless excitation branch with the Goldstone mode vanishes quickly for increasing magnetic field, whereas that associated with the gapped pseudo-Goldstone branch persists until the field-induced quantum phase transition to the uud phase \cite{Chi2024}. As shown in  Figs.~\ref{spectra_LET}(b) and (c), this also agrees well with our experimental observation. The diminished visibility of the roton dip at 0.5 T may stem from its shallowing as the magnetic field increases, making it more challenging to discern experimentally. Theoretically, it has been proposed that the roton modes in the pseudo-Goldstone branch arise due to a magnon-Higgs scattering mechanism \cite{Gao2024, Powalski2015}, while those in the gapless Goldstone branch result from energy level repulsion imposed by the former. This explains its disappearance under a magnetic field, as the pseudo-Goldstone branch shifts to higher energy [see Figure~\ref{spectra_LET}(c)].

Finally, we discuss the gapped continuum near 3 meV. Since \NaBaCoP ($\alpha = 0.6$) is much further from the Ising limit compared to \KSeCo ($\alpha = 0.07$), such excitations due to spin-flips are not present in the theories mentioned above. However, in a more recent theory for \KSeCo \cite{Xu2025}, such an excitation is indeed reproduced. Still, the transformation of the sharp, flat, dispersive-less spin-flip excitation in a pure Ising system into a continuum as the system deviates slightly from the Ising limit arises from the quantum fluctuations within the Wannier manifold induced by $J_{xy}$. If these excitations were due to closely spaced magnon modes, the energy difference would have to fall below our experimental resolution. The low-energy excitations have also been studied within the same theoretical framework, and the roton-like dip at the $\textrm{M}$ point is also predicted.

The XXZ triangular lattice model, in all its simplicity, is home to a plethora of very interesting and fundamental many-body quantum phenomena.
It is exceedingly rare to find a realization of such an important theoretical construct in a real material as exact, as what we have in \KSeCo. Moreover, it is very rare that a spin system where all the exotic physics stems from strong geometric frustration can be studied, and that on a quantitative level, with QMC. \KSeCo also stands out in that, by virtue of being in the strong-Ising limit, it gives us this unique opportunity.

\section*{Methods}

\textbf{Sample Synthesis.} Single crystals of \KSeCo are grown as in Ref. \cite{Zhong2020}. The crystal structure contains ABC-stacked two-dimensional triangular lattice planes of Co$^{2+}$ ions separated by K layers [Figure~\ref{fig:crys_str}]. The space group is $R$-3$\textrm{m}$, and the lattice parameters are $a=5.52$~\AA, and $c=18.52$~\AA~\cite{Wildner1994}.

\textbf{Inelastic neutron scattering.} High-resolution inelastic neutron scattering measurements were performed using the AMATERAS cold-neutron time-of-flight spectrometer \cite{AMATERAS} in J-PARC, Japan. The incident neutron energies used are $E_i$ =  $1.69$ and $7.73$ meV. The energy resolution at elastic scattering is $23$ and $240$ $\mu$eV, respectively.  The data were collected by rotating the sample by $160$ degrees with $1$ degree per step, and analyzed using the Horace package \cite{EWINGS2016-horace}. Inelastic neutron scattering measurements in a magnetic field were carried out using the LET cold-neutron time-of-flight spectrometer in ISIS facility, Rutherford Appleton Laboratory, United Kingdom. The incident neutron energy is fixed at $E_i$ = 2.15 meV. The spectrometer is operated in the high-flux mode, with chopper frequency 240/120 Hz, yielding an energy resolution of 40 $\mu$eV at elastic scattering. The data were obtained by rotating the sample by 90 degrees with 1 deg. per step. A single crystal sample of 425 mg and 1.23 grams were used for AMATERAS and LET, respectively. They are mounted on a copper holder and cooled using a $^3$He/$^4$He dilution refrigerator. At LET, a magnetic field is applied along the [0,0,1] direction using a vertical-field cryomagnet. In both cases the sample was oriented with [1,0,0] and [0,1,0] directions in the horizontal plane.

\textbf{Specific heat measurements.}
The specific heat measurements were performed using a relaxation method by a Quantum Design Physical Property Measurement System on a 0.05 mg single-crystal sample for 0.12$-$4 K with a dilution fridge insert and on a 3.95 mg one for 2$-$300 K. 

\textbf{LSWT  and QMC calculations.}
All LSWT calculations were performed using Sunny.jl (v0.7) \cite{SUNNY}. Worldline QMC calculations were performed using a modified directed-loop algorithm \cite{Kato2009, Motoyama2021} for system sizes of $36\times 36$ (thermodynamic properties) and $72\times 72$ (spin dynamics) with $N_{\tau}=200$ imaginary-time-steps. To verify the correctness of the QMC code, exact diagonalization calculations were performed on small clusters.

The QMC worm algorithm simulation enables the calculation of the imaginary-time two-point correlation function ${\cal G}^{zz}({\bf q}, \tau) = \langle S^z_{{\bf -q}}(\tau) S^z_{{\bf q}}(0)\rangle$, where $\tau$ is the imaginary time. This imaginary-time correlation function is related to the dynamical spin structure factor (DSSF) ${\cal S}^{zz}({\bf q}, \omega)$ via a Laplace transformation
\begin{equation}
        {\cal G}^{zz}({\bf q}, \tau) = \int_{-\infty}^{\infty} e^{-\omega\tau} {\cal S}^{zz}({\bf q}, \omega) d\omega.
        \label{Eq:LT}
\end{equation}

Since detailed balance dictates that ${\cal S}^{zz}_{-{\bf q}}(-\omega) = e^{-\beta \omega} {\cal S}^{zz}_{\bf q}(\omega)$, and we are interested in a problem with spatial inversion symmetry, ${\cal S}^{zz}_{-{\bf q}}(\omega)= {\cal S}^{zz}_{{\bf q}}(\omega)$, the integration in Eq.~\ref{Eq:LT} can be restricted to positive frequencies, 
\begin{equation}
    {\cal G}^{zz}({\bf q}, \tau) = \int_{0}^{\infty} K(\omega, \tau) {\cal S}^{zz}({\bf q}, \omega) d\omega,
    \label{Eq:G}
\end{equation}
with 
\begin{equation}
    K(\omega, \tau) = \left( e^{-\tau\omega} + e^{-(\beta - \tau)\omega}\right). 
\end{equation}

The inversion problem is non-trivial due the ill-posed nature of inverting~Eq.~\ref{Eq:G}, meaning that there is no unique solution. 
As a first approximation, we can use a saddle point approximation of Eq.~\ref{Eq:G}, which extracts the frequency of the lowest energy pole of ${\cal S}^{zz}({\bf q}, \omega)$. The frequency component of ${\cal G}^{zz}({\bf q}, \tau)$  is the sole survivor in the asymptotically long imaginary time:
\begin{equation}
    {\cal G}^{zz}({\bf q},\tau) \approx  A_{\bf q} \left( e^{-\tau\omega^*} + e^{-(\beta - \tau)\omega^*}\right) \ \ \ \text{as} \ \ \ \tau\rightarrow\beta
    \label{Eq:fitting}
\end{equation}
$A_{\bf q}$ and $\omega^*$ represent fitting parameters associated with the residue and frequency of the pole, respectively. These parameters are acquired through a non-linear fitting process \cite{Zhang2013}.

To attack the problem of inverting Eq.~\ref{Eq:G}, we employ the ``Differential Evolution Algorithm'' introduced and implemented in SmoQyDEAC.jl (v1.1.4) library~\cite{Nathan2022, Neuhaus2024}. After applying this procedure, the data is convoluted with the experimental resolution.

 

\section*{Data availability}
All data are available upon reasonable request to the corresponding author. The data collected at LET, STFC ISIS Neutron and Muon Source, is available at \url{https://doi.org/10.5286/ISIS.E.RB2320138}.

\section*{Code availability}
Sunny.jl (v0.7) and SmoQyDEAC.jl (v1.1.4) are publicly available on their respective GitHub repositories. The QMC code used for this study may be made available to qualified researchers upon reasonable request from the corresponding author.

\begin{acknowledgements}
We are grateful to M. E. Zhitomirsky for directing us to relevant prior literature and to A. Läuchli for illuminating discussions. We have incorporated new references to appropriately acknowledge these contributions. The authors are grateful to Hao Zhang for advice on using Sunny.jl. We thank James Neuhaus for the useful discussion in the setup of 
SmoQyDEAC.jl. This work was Supported by a MINT grant of the Swiss National Science Foundation. Data at J-PARC were collected in Experiment no. 2023B0161. Experiments at the ISIS Pulsed Neutron and Muon Source were supported by a beamtime allocation from the Science and Technology Facilities Council under proposal no. RB2320138. 
L. M. C. was supported by the U.S. Department of Energy, Office of Science, Basic Energy Sciences, Materials Science and Engineering Division.
\end{acknowledgements}

\section*{Author contributions}
The samples used in this study were synthesized by M.Z. and Z.Y. Thermodynamic measurements were performed by M.Z. and S.G.. Inelastic neutron scattering experiments were carried out by M.Z., V.R. and A.Z. with assistance from N.M., S.O. and C.B.  QMC calculations were performed by L.M.C and C.D.B. Y.K. developed and tested the QMC code. The manuscript was written by M.Z., L.M.C., A.Z. and C.D.B. with input from all authors.

\section*{Competing interests}
The authors declare no competing interests.

\bibliography{KSeCo}

\end{document}


\title{Supplemental Material for: ``Wannier states and spin supersolid physics in the triangular antiferromagnet \KSeCo"}

\author{M.~Zhu}
\thanks{These two authors contributed equally}
\address{Laboratory for Solid State Physics, ETH Z\"{u}rich, 8093 Z\"{u}rich, Switzerland}

\author{Leandro M. Chinellato}
\thanks{These two authors contributed equally;\\ Corresponding author:~ \url{lchinell@vols.utk.edu}}
\address{Department of Physics, The University of Tennessee, Knoxville, Tennessee 37996, USA}

\author{V. Romerio}
\address{Laboratory for Solid State Physics, ETH Z\"{u}rich, 8093 Z\"{u}rich, Switzerland}

\author{N. Murai}
\address{J-PARC Center, Japan Atomic Energy Agency, Tokai, Ibaraki 319-1195, Japan}

\author{S. Ohira-Kawamura}	
\address{J-PARC Center, Japan Atomic Energy Agency, Tokai, Ibaraki 319-1195, Japan}

\author{Christian Balz}
\address{ISIS Neutron and Muon Source, STFC Rutherford Appleton Laboratory, Didcot OX11 0QX, United Kingdom}
\address{Neutron Scattering Division, Oak Ridge National Laboratory, Oak Ridge, Tennessee 37831, USA}

\author{Z. Yan}
\address{Laboratory for Solid State Physics, ETH Z\"{u}rich, 8093 Z\"{u}rich, Switzerland}
\author{S. Gvasaliya}
\address{Laboratory for Solid State Physics, ETH Z\"{u}rich, 8093 Z\"{u}rich, Switzerland}

\author{Yasuyuki Kato}
\address{Department of Applied Physics, University of Fukui, Fukui 910-8507, Japan}

\author{C. D.~Batista}
\address{Department of Physics, The University of Tennessee, Knoxville, Tennessee 37996, USA}
\address{Quantum Condensed Matter Division and Shull-Wollan Center, Oak Ridge National Laboratory, Oak Ridge, Tennessee 37831, USA}

\author{A.~Zheludev}
\address{Laboratory for Solid State Physics, ETH Z\"{u}rich, 8093 Z\"{u}rich, Switzerland}

\begin{abstract}
	Contents: Supplementary Notes 1 – 4, Supplementary Figures 1 – 6.
\end{abstract}

\date{\today}
\maketitle
\section{Subtraction of nuclear Schottky contribution in heat capacity data}

\begin{figure}[h!]
    \centering
        \includegraphics[width=0.6\textwidth]{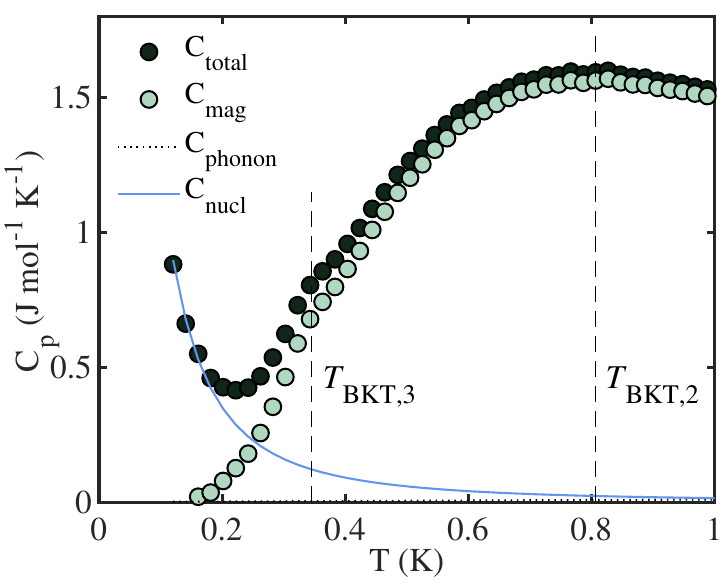}

    \caption{{\bf Nuclear Schottky contribution to the total heat capacity.}
    Temperature dependence of the total (solid symbols) and magnetic heat capacity (open symbols) of \KSeCo. The nuclear contribution (solid line) is estimated by fitting the Schottky anomaly of a two-level system as described in the text. The phonon contribution (dotted line) is estimated using the scaled heat capacity of a nonmagnetic isostructural compound K$_2$Mg(SeO$_3$)$_2$ as described in the main text. $T_{\text{BKT,2}}$ and $T_{\text{BKT,3}}$ denote temperatures of two BKT transitions. }
    \label{Fig:nuclear_heat_capacity}
\end{figure}

Supplementary Figure~\ref{Fig:nuclear_heat_capacity} shows the total and magnetic heat capacity of \KSeCo as a function of temperature, respectively. The nuclear contribution is estimated by fitting the low-temperature upturn by the Schottky anomaly of a two-level system
\begin{equation}
C_{\mathrm{nucl}} = R(\frac{\Delta}{T})^2\frac{\mathrm{exp}(\Delta/T)}{[1+\mathrm{exp}(\Delta/T)]^2}
\end{equation}
where $\Delta$ is the energy between the two levels. The best fit is obtained for $\Delta = 0.084(1)$ K. 
The weak heat capacity anomaly corresponding to the lowest temperature BKT transition at $T_{\text{BKT,3}}$ is evident in the raw data. Its characteristic temperature is barely affected by the subtraction process.

A similar upturn in the heat capacity is reported in the bilayer system K$_2$Co$_2$(SeO$_3$)$_3$ in zero field (denoted as T$_{c3}$ in \cite{YingFu2025}), which extends up to about 3 T into the 1/3-plateau phase and is attricuted to be the transition to the ``Y"-phase. However, for K$_2$Co(SeO$_3$)$_2$ in this study, the magnetic phase diagram (Fig.~2(b) of \cite{Zhu2024}) shows that this upturn of heat capacity is clearly a separate feature away from the quantum critical point into the 1/3-plateau phase, and the phase transition into the ``Y" phase occurs already around T$_{\text{BKT,3}}$ (noted as T$_c$ in \cite{Zhu2024}).

\section{Dispersion along the $L$ direction}
\begin{figure}[h!]
    \centering
        \includegraphics[width=0.6\textwidth]{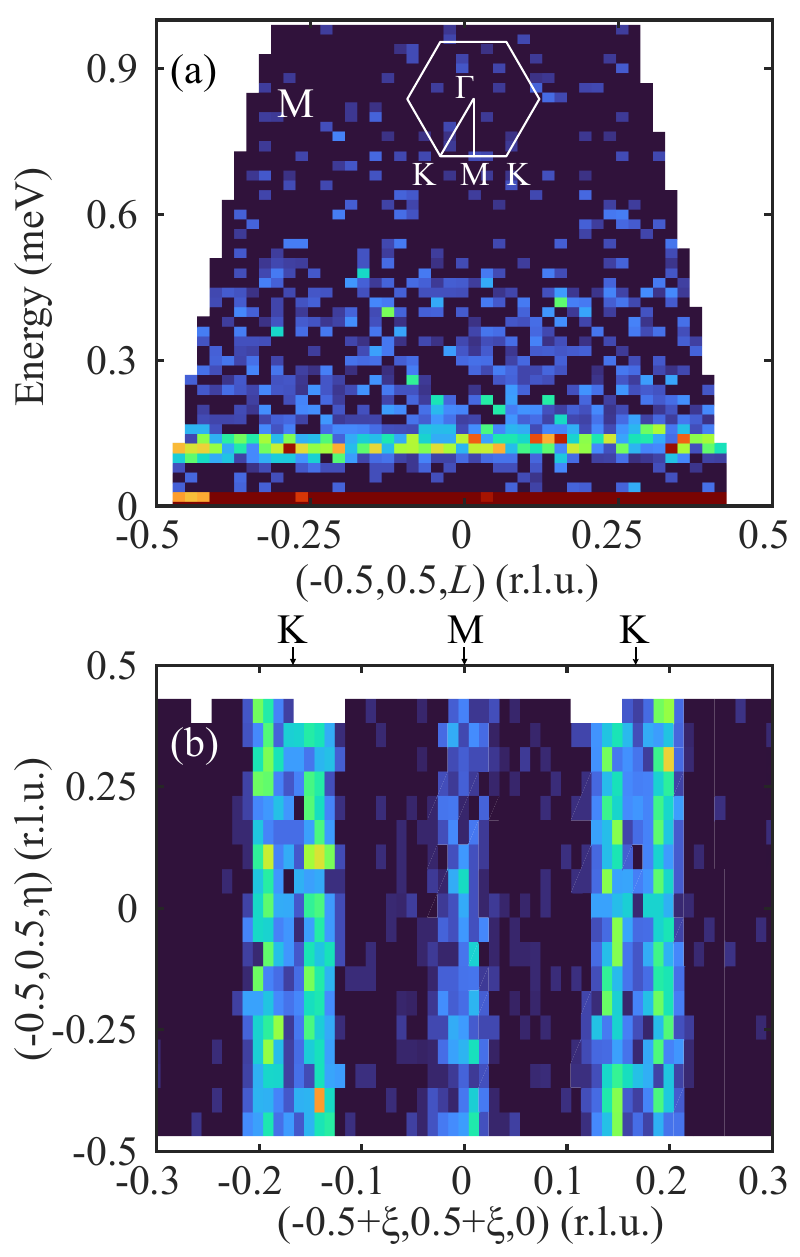}

    \caption{{\bf $L$ dependence of the magnetic excitations.}
    (a) False-color plot of the magnetic excitation spectrum of \KSeCo at the $M$ point as a function of energy along the (0,0,$l$) direction, measured by inelastic neutron scattering in zero field and $T$ = 0.07 K at AMATERAS with $E_i$ = 1.69 meV. The data are integrated over $\pm$0.03 r.l.u. along two orthogonal directions in the ($h,k$,0) plane, respectively. (b) Constant energy cut of the magnetic excitation spectrum as a function of ($h,h$,0) (along the path $K-M-K$ in the reciprocal space) and (0,0,$l$) direction, with energy integrated from 0.09 to 0.13 meV corresponding to the lower boundary of the continuum at the $M$ point. The integration range in the orthogonal direction of the ($h,k$,0) plane is $\pm$0.03 r.l.u.. White Hexagon represents the boundary of the Brillouin zone. The data are not averaged over equivalent paths in the reciprocal space. }
    \label{Fig:L_dependence}
\end{figure}

Supplemental Figure~\ref{Fig:L_dependence} shows the representative cuts of the magnetic excitation spectrum of \KSeCo across the $M$ point of the Brillouin zone along the $L$ direction, measured in zero magnetic field and $T=$ 70 mK. The flat band in (a) corresponds to the lower-boundary of the continuum of excitations at the $M$ point and its energy exhibits no $L$ dependence. (b) is a constant-energy cut as a function of ($h,h$,0) and (0,0,$l$), with energy integrated from 0.09 to 0.13 meV (at the lower boundary of the continuum at the $M$ point). The rod-like features again affirm that the excitations are dispersionless along the $L$ direction.

\section{Reciprocal paths}
\begin{figure*}[h!t]
	\includegraphics[width=0.98\textwidth]{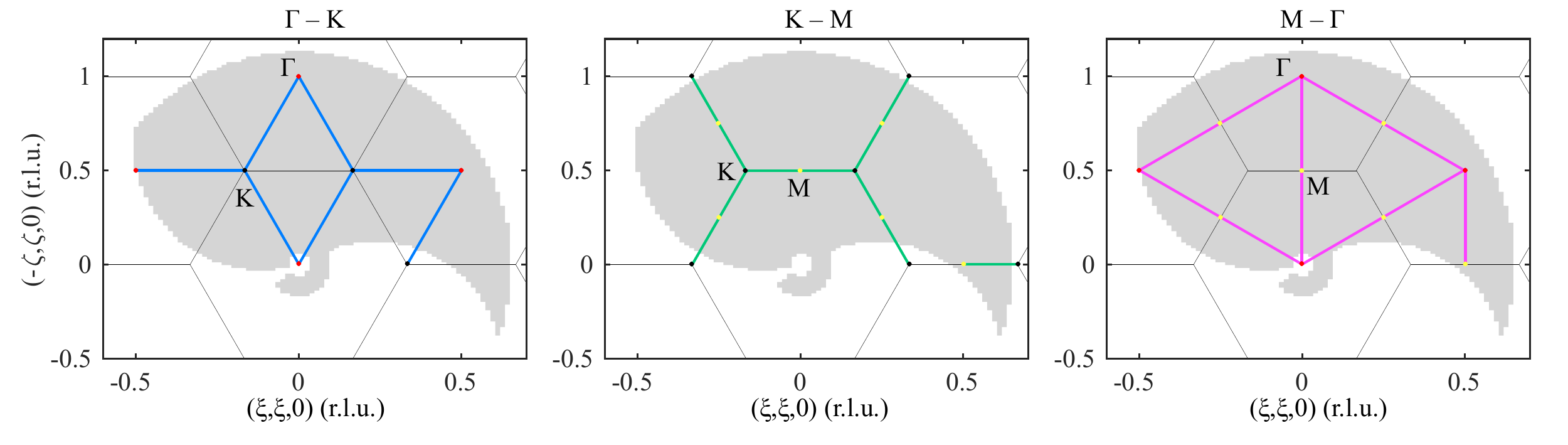}
	\caption{\textbf{Coverage of reciprocal space in AMATERAS with $E_i$ = 1.69 meV and paths of high-symmetry directions.} Schematic diagram showing the equivalent reciprocal space paths (a) $\Gamma$-$\textrm{K}$, (b) $\textrm{K}$-$\textrm{M}$, and (c) $\textrm{M}$-$\Gamma$, used to plot the averaged excitation spectrum in Figure 5 of the main text. 
     The shaded area is the reciprocal space covered by the detector bank at AMATERAS with $E_i$ = 1.69 meV. Black hexagons represent the Brillouin zone boundaries. }\label{AMA_Q_range}
\end{figure*}

Supplementary Figure~\ref{AMA_Q_range}$-$\ref{AMA_highEi_Q_range} illustrate the reciprocal paths used to plot the averaged excitation spectrum in Fig.~5, 7, and 10 of the main text.

\section{Dynamical spin structure factor computed by LSWT}
\label{Ap:LSWT}

Supplementary Figure~\ref{Fig:LSWT_dssf} shows the total (left) and ${\cal S}^{zz}$ component (right) of the dynamical spin structure factor computed using linear spin-wave theory.

\newpage

\begin{figure*}[h!t]
	\includegraphics[width=0.98\textwidth]{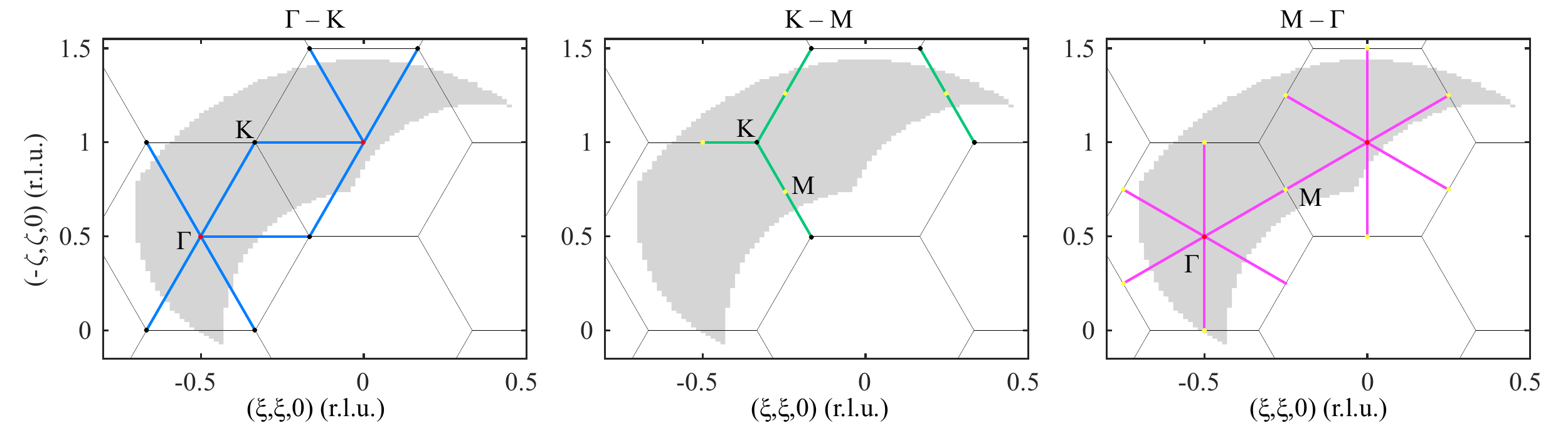}
	\caption{\textbf{Coverage of reciprocal space in LET with $E_i$ = 2.15 meV and paths of high-symmetry directions.} Schematic diagram showing the equivalent reciprocal space paths (a) $\Gamma$-$\textrm{K}$, (b) $\textrm{K}$-$\textrm{M}$, and (c) $\textrm{M}$-$\Gamma$, used to plot the averaged excitation spectrum in Figure 12 of the main text. 
     Shaded area is the reciprocal space covered by the detector bank at LET with $E_i$ = 2.15 meV. Black hexagons represent the Brillouin zone boundaries.  }
\end{figure*}

\begin{figure*}[h!t]
	\includegraphics[width=0.98\textwidth]{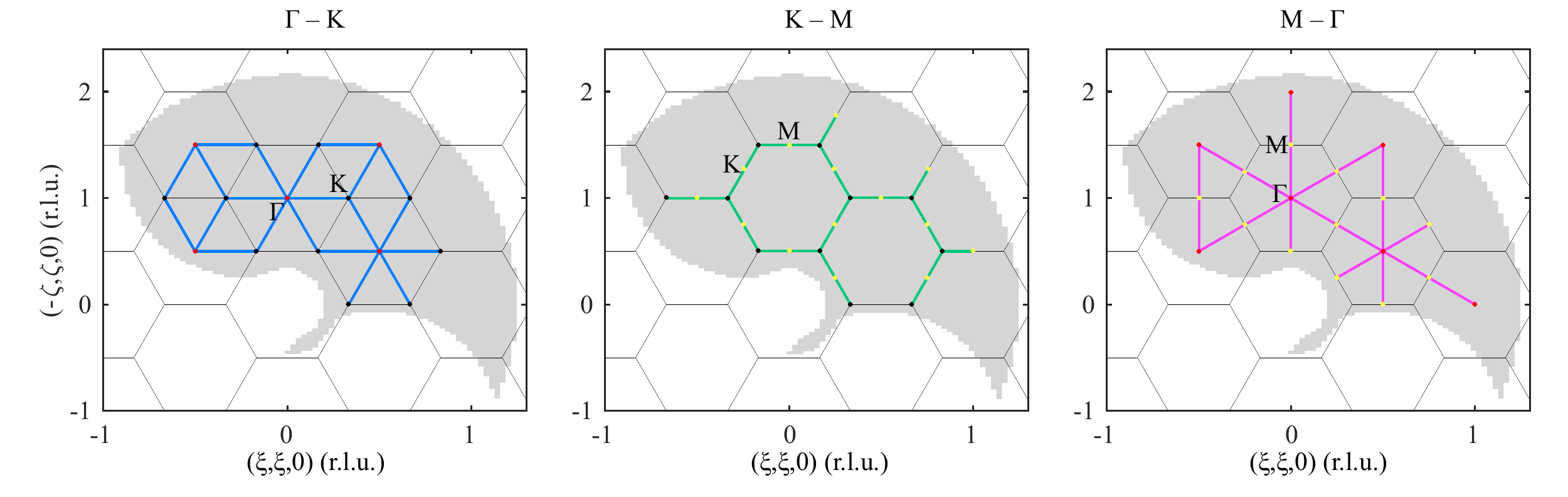}
	\caption{\textbf{Coverage of reciprocal space in AMATERAS with $E_i$ = 7.73 meV and paths of high-symmetry directions.} Schematic diagram showing the equivalent reciprocal space paths (a) $\Gamma$-$\textrm{K}$, (b) $\textrm{K}$-$\textrm{M}$, and (c) $\textrm{M}$-$\Gamma$, used to plot the averaged excitation spectrum in Figure 16 of the main text.  
    The shaded area is the reciprocal space covered by the detector bank at AMATERAS with $E_i$ = 7.73 meV. Black hexagons represent the Brillouin zone boundaries.  }\label{AMA_highEi_Q_range}
\end{figure*}

\begin{figure}[h!]
    \centering
        \includegraphics[width=0.6\textwidth]{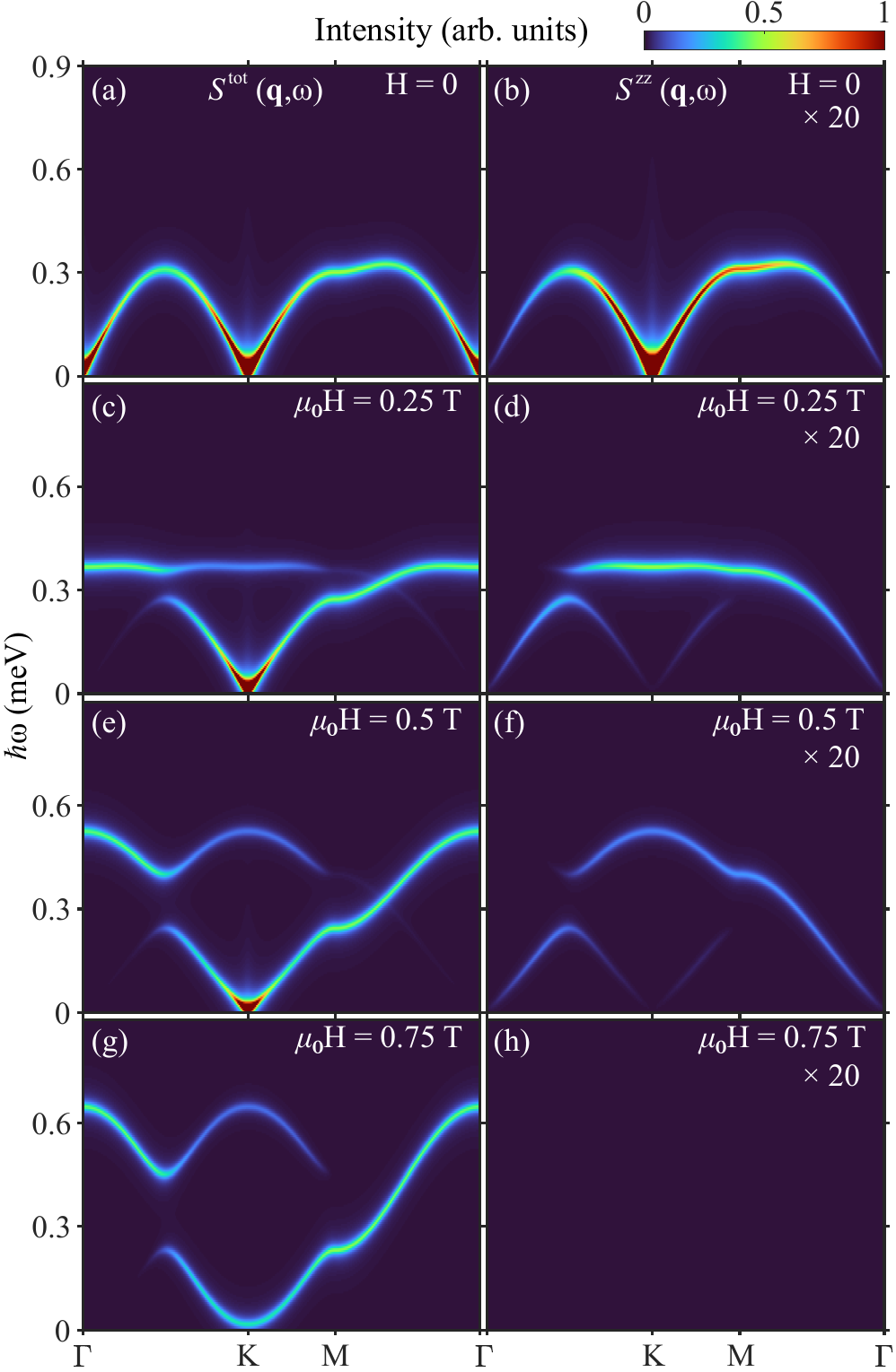}

    \caption{{\bf Evolution of the DSSF in the supersolid phase under a magnetic field, computed with LSWT.}
    Total (left column) and ${\cal S}^{zz}$ component (right column) of the dynamical spin structure factor computed by linear spin wave theory for (a),(b) $H=0$ T, (c),(d) $\mu_0 H=0.25$ T and (e),(f) $\mu_0 H=0.5$ T and (g),(h) $\mu_0 H=0.75$ T. The FWHM used in the Dirac delta functions corresponds to the experimental resolution. Note the $\times 20$ scale factor in the intensity of ${\cal S}^{zz}$. }
    \label{Fig:LSWT_dssf}
\end{figure}

\clearpage
\bibliography{KSeCo}